\documentclass[aps,twocolumn,reprint]{revtex4-1}
\usepackage{graphics}
\usepackage{subfigure}
\usepackage{epsfig}
\usepackage{epsf,epic}
\usepackage{color}
\usepackage{marvosym }
\usepackage{subfigure}
\usepackage{siunitx}
\usepackage{amsmath}
\usepackage{amssymb}
\usepackage{amsfonts}
\usepackage{wrapfig}
\usepackage{pstricks}
\usepackage{multirow}
\usepackage{pst-node}
\usepackage{extdash}
\usepackage{braket}
\usepackage{bm}
\usepackage{dcolumn}
\newcommand{\etal}{\textit{et al\ }}

\newcommand{\half}{\frac{1}{2}}
\newcommand{\quarter}{\frac{1}{4}}
\begin{document}
\title{Calculated phonon modes, infrared and Raman spectra in orthorhombic $\alpha$-MoO$_{3}$ and monolayer MoO$_3$}
\author{Amol Ratnaparkhe}
\author{Santosh Kumar Radha}
\author{Walter R. L. Lambrecht}
\email{Corresponding author: walter.lambrecht@case.edu}
\affiliation{Department of Physics, Case Western Reserve University, 10900 Euclid Avenue, Cleveland, Ohio 44106-7079, USA}
\begin{abstract}
  Orthorhombic   $\alpha$-MoO$_3$ is a layered oxide with various applications and with excellent potential
  to be  exfoliated as a 2D ultra-thin film or monolayer.  
  In this paper, we present a first-principles computational
  study of its vibrational properties. Our
  focus is on the zone center modes which can be measured by a combination
  of infrared and Raman spectroscopy. The polarization dependent spectra
  are simulated.  Calculations are also performed for a monolayer form
  in which ``double layers'' of Mo$_2$O$_6$ which are weakly van der Waals
  bonded in the $\alpha$-structure are isolated.  Shift in phonon frequencies
  are analyzed. 
  \end{abstract}
\maketitle
\section{Introduction}\label{intro}
MoO$_3$ is a layered transition metal oxide which has found various applications
in chemical sensing,\cite{Rahmani2010,Balendhran2013} batteries,\cite{Li2016}
catalysis,\cite{Voiry2013}
and as hole-extraction layer in organic
photovoltaic cells.\cite{Kroger2009} The latter application is based on its very high electron affinity,
meaning that the conduction band energy levels lie deep
below the vacuum level and can thus line up with the highest occupied molecular
orbitals (HOMO) in organic dye molecules used in organic photovoltaics. Being native n-type it can then replenish
the holes in the organic dye caused by photo absorption. 
Recently, MoO$_3$ thin films  were also found to have high dielectric
constant and were used as the gate oxide in thin film transistors.\cite{Holler2020}

From a more fundamental science point of view, MoO$_3$ is an excellent  candidate oxide for exfoliation to mono- or few-layer ultrathin films.\cite{Balendhran2013am} In that sense
it is comparable to V$_2$O$_5$ because in both cases, the transition metal
is in its highest possible valence state and they both form layered crystals
with weak van der Waals interactions between the neutral layers.
The band structure consists of filled oxygen orbital derived
valence bands and empty metal d-states. The layered structure is also
the starting point for  intercalation of alkali metal elements between
the layers, leading to so-called bronze structures which have potential applications in batteries.

Mechanical exfoliation
was recently successfully applied to V$_2$O$_5$  and showed extremely
anisotropic behavior or the in-plane electron transport related to 1D chain like
elements of the structure.\cite{Sucharitakul2017}
It was also  predicted that some of the vibrational
modes would show large blue shifts when going from 3D to the monolayer form.\cite{Bhandari14}
Hence the vibrational properties of MoO$_3$ are also of significant interest.

While there have been prior Raman and infrared studies,\cite{Mestl1994,Py81,Seguin1995,Eda91}
a full first-principles analysis of the
vibrational properties and polarization dependent Raman spectra on
single crystals or thin films have not yet been reported. Among other this
may assist in the characterization of ultra-thin layers or nanoflakes of
MoO$_3$.
Here we present a first-principles calculation of the phonons in
$\alpha$-MoO$_3$ including simulations of the Raman and infrared spectra and
study the changes in phonon spectra between bulk $\alpha$-MoO$_3$ and monolayer
MoO$_3$.

\section{Computational Methods}\label{compmeth}
The calculations are done using Density Functional Perturbation Theory
(DFPT)\cite{Gonze1,Gonze2}
using the plane-wave pseudopotential method as implemented
in the ABINIT \cite{abinit,abinit20} and Quantum Espresso codes.\cite{Giannozzi_2009}
Specifically, with the ABINIT code we  choose
the Hartwigsen-Goedecker-Hutter pseudopotentials \cite{Hartwigsen98}
and the local density approximation (LDA). The performance of LDA and
generalized gradient correction (GGA) in the Perdew-Burke-Ernzerhof (PBE)
\cite{PBE} approximation as well as other exchange correlation functionals
were studied systematically for phonons in
Ref. \onlinecite{He14} and find generally LDA to be closer to experiments. 
The energy cutoff used in these calculations  is $160$ Rydberg, which was tested first to give converged results. 
For the Brillouin zone integration or charge densities and total energy 
a $4\times4\times4$ \textbf{k}-point mesh is used.   
Phonon calculations are done at the $\Gamma$-point. These are sufficient to 
determine the infrared absorption and reflection (IR) spectra as well as  the
Raman spectra assuming momentum conservation and using that visible
and infrared light has negligible momentum compared to the Brillouin zone
size. The Raman spectra are calculated
using the approach of Veithen \etal\cite{Veithen}.
Various associated quantities, such as the Born effective charges,
electronic dielectric susceptibilities and oscillator and strength and
Raman tensors can all be obtained from second or third order
derivatives of the total energy versus atomic displacements
and homogeneous electric field
components within DFPT. 

To further test these results, we also used the Quantum Expresso code
with projector augmented wave (PAW)\cite{Blochl94}
pseudopotentials generated by Dal Corso\cite{DalCorso}
and using the GGA-PBEsol exchange correlation functional.\cite{PBEsol}
In this case a 120 Ry cutoff was used for kinetic energy and 480 Ry for
charge densities. These calculations were used for the orthorhombic 
and the monolayer structures. 
\section{Results}\label{results}
\subsection{Crystal structure and group theoretical analysis}\label{group theory}
The space group of $\alpha$-MoO$_3$ is $Pmcn$  number 62 (or $D_{2h}^{16}$. Note that the standard setting of the International
Tables for Crystallography is $Pnma$ but then the normal mirror plane, labeled
$m$,
is perpendicular to ${\bf b}$ whereas ours is perpendicular to ${\bf a}$.
In the symmetrized version of the Materials Project coordinates,\cite{MP} corresponding
to $Pnma$ the direction normal to the layers is the $a$ direction, which
is our $c$. Hence the double glide mirror plane, labeled $n$, is perpendicular to ${\bf c}$
in our case. The point group is $D_{2h}$.
The unit cell of the   structure contains 16 atoms,
4 Mo and 12 O atoms, which belong to three different types. 

The optimized lattice constant  within LDA and reduced coordinates 
are given in Table \ref{tabstruc}. We also give the volume of the cell $V$
in this table and compare  our lattice constants with the experimental
ones by Seguin \etal \cite{Seguin1995} and with the ones from
Materials Project \cite{MP} optimized in the generalized gradient approximation (GGA)
in the Perdew-Burke-Ernzerhof (PBE) parameterization \cite{PBE} and
with the PBEsol results calculated with Quantum Espresso. 
Note that Seguin \etal \cite{Seguin1995}
used yet another setting $Pbnm$ where the largest lattice constant
normal to the layers is the $b$ direction but we converted their results
to our present setting of the space group. 
We may note that our
lattice volume is slightly underestimated  compared to the experiment
while the GGA PBE value is 6 \% overestimated.  However, we should also
note that this overestimate is mostly stemming from the $c$-lattice constant
overestimate, which is 4 \%,  and only about 2 \% from the $a$ lattice constant.
The $c$ lattice direction is perpendicular to the layers and thus
most sensitive to the weak van der Waals interactions. The good agreement
for this in LDA may be somewhat spurious and does not indicate that
LDA should always perform well on such interlayer interactions but
is useful here.  Our $b/a$ ratio at 1.062 is intermediate between the experimental
value of 1.072 and the PBE value of 1.055. Our $c/a$ ratio at 3.708 is smaller than the experimental
value of 3.748 and the Materials Project\cite{MP}  value of 3.855.
In the PBEsol functional, we find an even larger overestimate of
the $c$-lattice constant with a $c/a=4.595$.
This indicates the difficulty of these exchange-correlation functionals
to treat weakly van der Waals bonded layer interactions.  The distance
between the center of the bilayers equals $c/2$ and thus $c$ gives directly
an indication of the separation of the bilayers and is 22 \%
overestimated by PBEsol and 4 \% by PBE but underestimated by -0.5 \%
by LDA. 

\begin{table}
  \caption{Reduced coordinates and lattice constants in
    $\alpha$-MoO$_3$ in the $Pnma$ space group. \label{tabstruc}}
  \begin{ruledtabular}
    \begin{tabular}{lcccc}
      atom & Wyckoff & $x$ & $y$ & $z$ \\ \hline
      Mo   &    4c     & 0.25   &  0.91613   &  0.60637  \\
      O$^{(1)}$  &  4c     & 0.25   &  0.46959    &  0.58859   \\
      O$^{(2)}$  & 4c      & 0.25   &  0.95868    &  0.72917 \\
      O$^{(3)}$  & 4c      & 0.25   &  0.50056    &  0.93527   \\
      \hline
         & $a$ (\AA)& $b$ (\AA)  & $c$ (\AA) & $V$ (\AA$^3$)\\ \hline 
      Calc. (LDA) & 3.7217 & 3.9510  & 13.7916 & 202.798  \\
 MP (PBE)\cite{MP}    & 3.761   & 3.969 & 14.425 & 215.328\\
      Calc. (PBEsol) & 3.682   & 3.860    &  16.919       & 240.462 \\
      Expt. \cite{Seguin1995} & 3.6964 & 3.9628&  13.855  & 202.949 \\
         \end{tabular}
  \end{ruledtabular}
\end{table}

\begin{figure}
  \caption{Crystal structure of $\alpha$-Mo in the $Pmcn$ space group
    setting. Viewed in terms of pyramidal or distorted octahedral units.\label{figstruc}}
  \includegraphics[width=4.5cm]{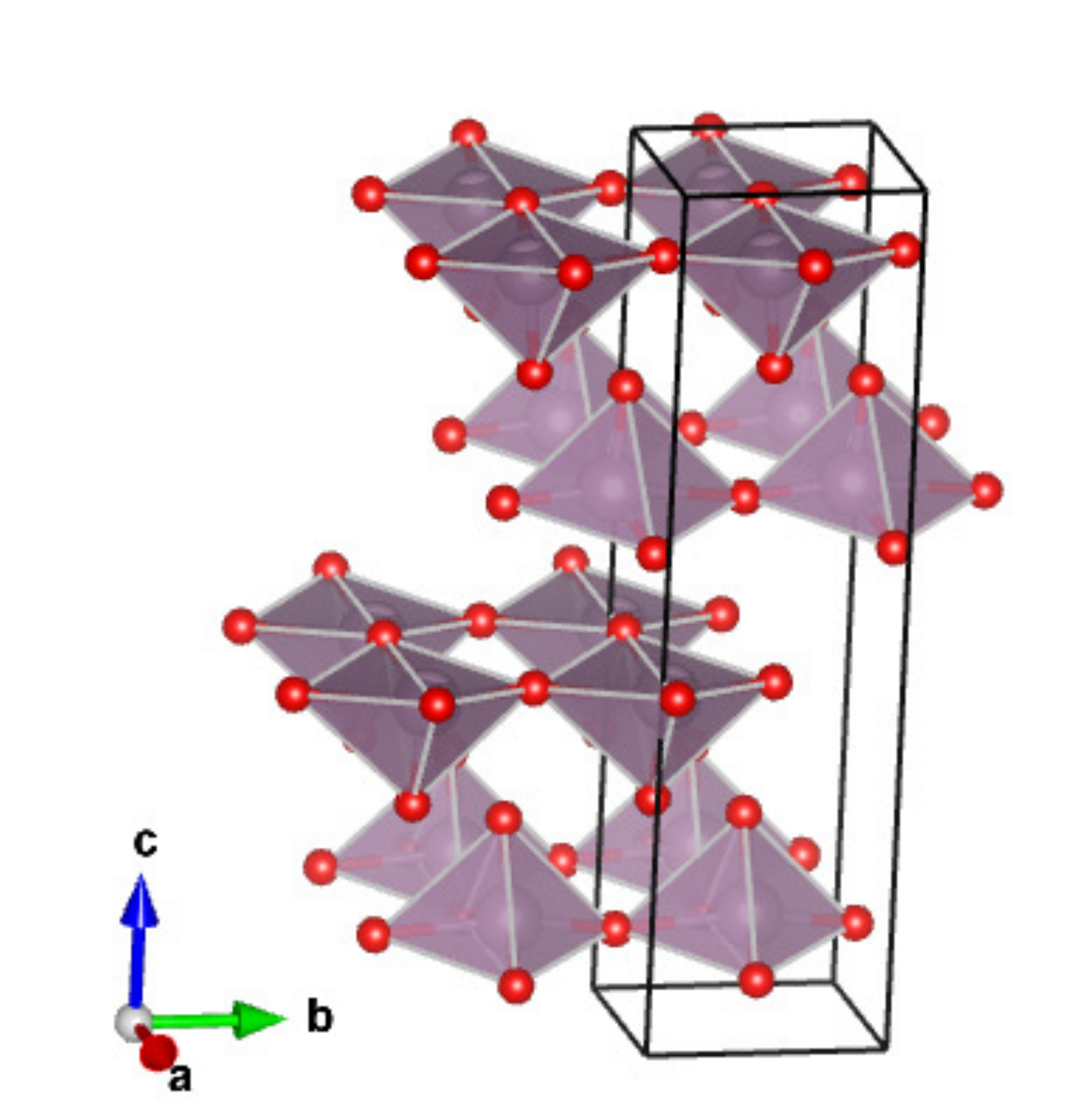}
  \includegraphics[width=4.cm]{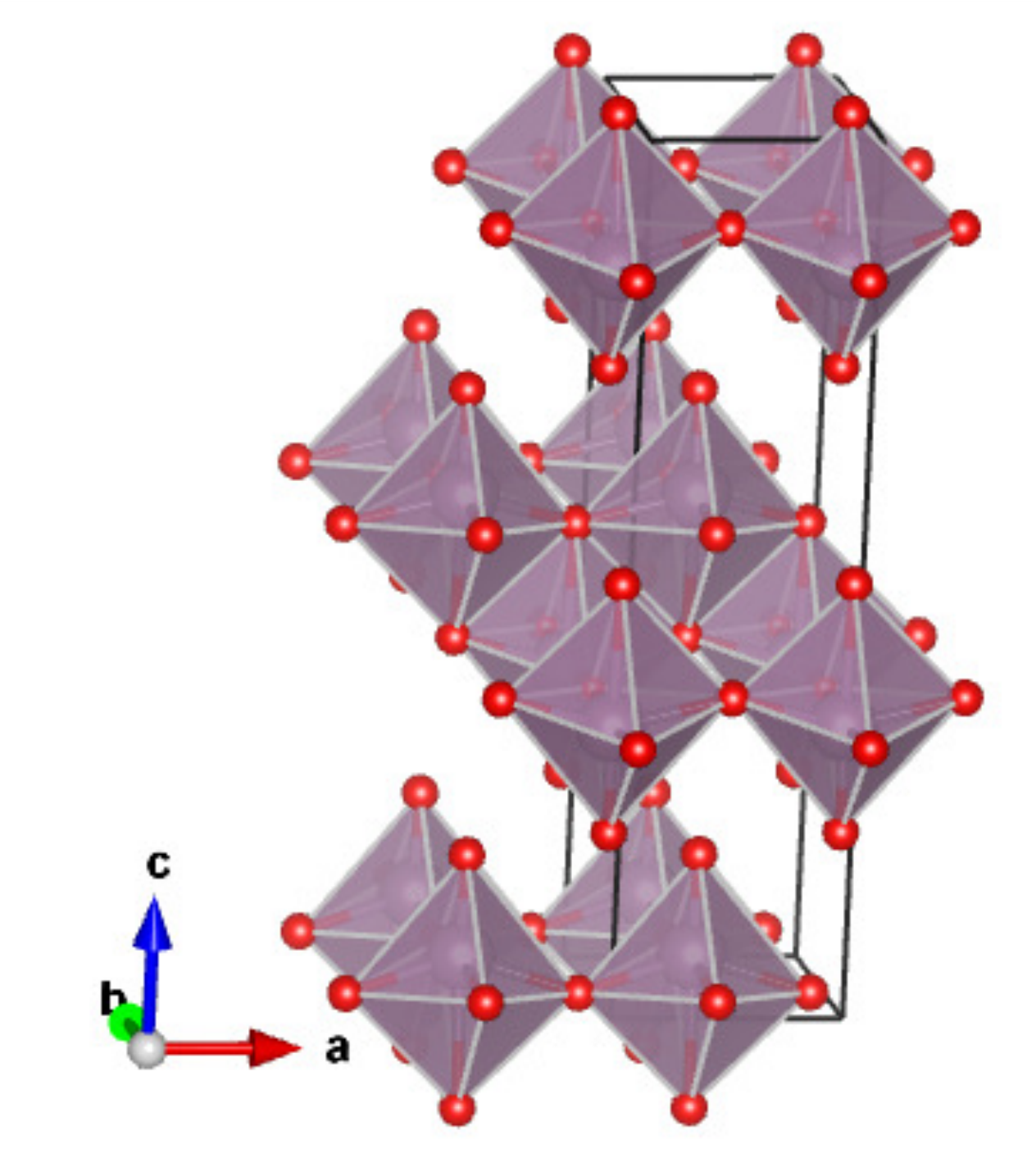}
\end{figure}

The crystal structure is shown in Fig. \ref{figstruc}.
Note that the O$^{(2)}$ is bonded to a single Mo and has a short bond
of only 1.702 \AA. O$^{(1)}$ is bonded to two Mo in a bridge configuration
along the $b$ direction with alternating bond lengths of 1.781  \AA\ and
2.218 \AA. O$^{(3)}$ is bonded to two Mo along the $a$ direction each at
1.978 \AA\ but also to another Mo at a larger distance of 2.386 \AA\ in the $c$-direction.
When this last long bond is ignored the structure can be
described in terms of slightly distorted square pyramids which are corner
sharing in the $ab$-plane. Two adjacent layers of such pyramids face each
other via their flat faces and form a double layer
with the short Mo-O$^{(2)}$ bonds facing outward. These double layers are
weakly van der Waals bonded. When the longer bond of 2.386 is included slightly
in the coordination polyhedron, the structure can be viewed as consisting
of distorted octahedra which share edges with the lower octahedron 
in the $a$ direction and corners in the $a$ and $b$-directions.
Along the $c$ axis double layers are stacked with a van der Waals gap
formed between the O$^{(2)}$ single bonded oxygens.

\begin{table}
  \caption{Character table of $D_{2h}$.\label{chard2h}}
  \begin{ruledtabular}
  \begin{tabular}{c|cccccccc|c}
    irrep           & $E$ & $I$ &  $C_{2y}$ & $m_{xz}$ & $C_{2z}$ & $m_{xy}$ & $C_{2x}$ & $m_{yz}$& basis function \\ \hline
    $A_{g}$   & 1  &     1         &     1      &   1       &    1       &       1     &     1       &      1       &            $x^{2}, y^{2}, z^{2}$ \\
    $B_{1g}$   & 1  &     1         &   $-1$   &  $-1$   &    1       &       1     &   $-1$   &  $-1$      &            $xy$ \\
    $B_{2g}$   & 1  &     1         &      1     &     1     &  $-1$    &   $-1$    &   $-1$   &  $-1$      &            $xz$ \\
    $B_{3g}$   & 1  &     1         &   $-1$   &  $-1$   & $-1$     &   $-1$   &       1      &     1       &            $yz$ \\
    $A_{u}$   & 1  &   $-1$      &      1     &  $-1$   &    1       &   $-1$   &       1      &  $-1$     &            $xyz$\\
    $B_{1u}$   & 1  &   $-1$      &   $-1$   &    1      &    1       &   $-1$   &   $-1$     &     1       &            $z$\\
    $B_{2u}$   & 1  &   $-1$      &     1      &  $-1$   &  $-1$    &      1     &   $-1$     &     1       &            $y$\\
    $B_{3u}$   & 1  &   $-1$      &   $-1$   &     1     &   $-1$   &      1     &      1       &  $-1$     &            $x$\\
  \end{tabular}
  \end{ruledtabular}
\end{table}
The character table of the point group $D_{2h}$ is given in
Table \ref{chard2h}. Further information on the space group
operations is given in Appendix. Excluding the translations along the three directions, $x$, $y$ and $z$, the vibrational modes are distributed
over the irreducible representations as
\begin{equation}
\begin{aligned}
  \Gamma^{vib}= & 8 A_{g} \oplus 4 B_{1g}\oplus 4 B_{2g}\oplus  8 B_{3g} \oplus 4 A_{u}\\
                            & \oplus  7 B_{1u}\oplus  7 B_{2u} \oplus 3 B_{3u} .
  \end{aligned}
\end{equation}
Of these modes, $A_{1u}$ modes are silent, the $B_{1u}$, $B_{2u}$ and $B_{3u}$ are infrared active
and show a LO-TO splitting for electric fields along $z=c$, $y=b$, $x=a$.
while $A_{g}$, $B_{1g}$, $B_{2g}$ and $B_{3g}$ are Raman active.
More precisely, 
the Raman tensors are of the form
\begin{eqnarray}
  \left(\begin{array}{ccc}a&.&.\\.&b&.\\.&.&c\\
  \end{array}\right), &\quad& \hbox{\rm for $A_{g}$,} \nonumber \\
   \left(\begin{array}{ccc}.&d&.\\d&.&.\\.&.&.\\
  \end{array}\right),  &\quad& \hbox{\rm for $B_{1g}$,} \nonumber \\
  \left(\begin{array}{ccc}.&.&e\\.&.&.\\e&.&.\\
  \end{array}\right),  &\quad& \hbox{\rm for $B_{2g}$,} \nonumber \\
  \left(\begin{array}{ccc}.&.&.\\.&.&f\\.&f&.\\
  \end{array}\right),  &\quad& \hbox{\rm for $B_{3g}$.} \nonumber \\ \label{eqtensors}
\end{eqnarray}

For the monolayer structure, we consider only one  ``double layer''
per cell and stack these directly on top of each other with  large
spacing in the $c$-direction.
The space group then becomes $P2_1/m$, which is, in principle,
monoclinic. In fact, a  structure with this space group is described in
Materials Project\cite{MP} but with shorter interlayer distances. The layers then
slide over each other and the structure becomes monoclinic with a $\alpha$-angle
different from 90$^\circ$. However, we space these layers much further
to effectively study a single isolated monolayer and hence there is no
driving force for this monoclinic distortion.
The crystal structure in this case has $a=3.7217$ \AA, $b=3.951$ \AA,
and $c=27.5832$ \AA\ as optimized in LDA. In GGA-PBE, they are
$a=3.7610$, $b=3.9693$, $c=28.8490$ \AA.
The point group in this case is $C_{2h}$ with
the two-fold (screw) axis along $x$, a mirror-plane $m_x$ and the inversion
center.  The irreducible representation  are given in the character table
\ref{tabcharc2h}
and their relation to those in the $D_{2h}$ group is also given. 

\begin{table}
  \caption{Character table for $C_{2h}$ relevant to the monolayer MoO$_3$
  and relation to parent irreducible representations of $D_{2h}$.\label{tabcharc2h}}
  \begin{ruledtabular}
    \begin{tabular}{c|rrrr|c|c}
    irrrep  & $E$ & $C_{2x}$ & $m_x$ & $i$ & basis functions & $D_{2h}$ parent\\ \hline
    $A_g$ &  1  &    1     &   1   &  1  & $x^2$, $y^2$, $z^2$, $yz$ & $A_g$, $B_{3g}$ \\
    $A_u$ &   1 &    1     & $-1$  &  $-1$ & $x$ & $B_{3u}$\\
    $B_g$ &   1 &   $-1$   & $-1$   &  1    & $xy$, $xz$ & $B_{1g}$,  $B_{2g}$ \\
    $B_u$ &   1 &   $-1$   & 1      & $-1$  & $y$, $z$  & $B_{2u}$, $B_{1u}$ \\ \hline
    \end{tabular}
  \end{ruledtabular}
\end{table}

\subsection{Phonon frequencies and related results.}\label{phonmodes}
The phonon frequencies at $\Gamma$ are given in Table \ref{tabmodes} both in LDA (calculated with ABINIT)
and  in PBESOL (calculated with Quantum Espresso) and compared with experimental values.
The PBEsol phonon calculation actually used the PBE optimized lattice constants from MP\cite{MP}
but with re-optimized internal coordinates.  The reason for doing this,
is that the $c$-lattice constant in PBEsol is clearly
overestimated as mentioned earlier. 
Corresponding to the light polarized along \textbf{z}, \textbf{x} or \textbf{y}, the LO-TO splittings are observed for
$B_{1u}$, $B_{3u}$ and $B_{2u}$ modes, respectively. From Table \ref{tabmodes},
we can observe that the splittings are significantly smaller for the lower frequency
modes compared to the higher frequency modes. This is because only the high frequency modes
have significant bond stretch dipolar character. 
The larger LO-TO splittings are also correlated with stronger oscillator
strengths for infrared absorption.
\begin{table*}
  \caption{Phonon frequencies of $\alpha$-MoO$_3$ in cm$^{-1}$. Experimental values
    from Seguin \etal \cite{Seguin1995}.  \label{tabmodes}}
  \begin{ruledtabular}
    \begin{tabular}{drr|drr|drr|drr|drr|drr} 
      \multicolumn{3}{c|}{$B_{1uT}$} & \multicolumn{3}{c|}{$B_{1uL}$} & \multicolumn{3}{c|}{$B_{2uT}$} & \multicolumn{3}{c|}{$B_{2uL}$} & \multicolumn{3}{c|}{$B_{3uT}$}& \multicolumn{3}{c}{$B_{3uL}$}    \\
      \multicolumn{1}{c}{LDA} & \multicolumn{1}{c}{PBEsol} & expt & \multicolumn{1}{c}{LDA}& \multicolumn{1}{c}{PBEsol} & expt &
        \multicolumn{1}{c}{LDA} & \multicolumn{1}{c}{PBEsol} & expt & \multicolumn{1}{c}{LDA}& \multicolumn{1}{c}{PBEsol} & expt &
           \multicolumn{1}{c}{LDA} & \multicolumn{1}{c}{PBEsol} & expt & \multicolumn{1}{c}{LDA}& \multicolumn{1}{c}{PBEsol} & expt \\ \hline
     53.24 & 21.44 & 53  & 53.28 & 21.45 & 53 & 46.88& 34.21 &44  & 46.98&34.22 & 44 & 174.14 & 178.87 & 191 & 176.97 & 179.29 & 191 \\   
    241.45 & 247.52 &260 & 243.10 & 248.78 & 260 & 213.97 & 231.24 & 228 & 214.00 &231.67 &228 & 244.22 &234.86 &268& 322.23& 336.30 & 343  \\  
     303.31 & 339.02 & 353 & 312.88 & 342.76 & 363&  301.57 & 339.47 &348 & 321.22 & 344.22 &352 & 535.38 & 477.63 & 545   & 799.16 & 785.11 & 851 \\ 
     333.42 & 348.93 &374 & 334.62& 349.09& 380\footnote{Calculated Py and Mashke\cite{Py81}} & 327.98& 348.49 & 363  & 351.13 &357.70 &390&     & &           & & &              \\  
      428.85 &429.52 & 441 &477.10 & 476.65  &505&  476.96 & 477.74 & 500 & 490.32 &491.40 & 525 && &             &  &   &          \\  
      725.40&766.50 &814   & 737.92 & 773.37 &825 &  732.49&774.77  & 818  & 906.12 & 942.66 & 974           &&&  & &              \\ 
      909.44 &983.48 &962 & 959.67 & 1032.23& 1010 & 948.17 & 1025.57 & 1002 & 948.17 & 1026.34 & 1002             &&& &&              
        \end{tabular}
\begin{tabular}{drr|dd|drr|drr|drr}    
     \multicolumn{3}{c|}{$A_{g}$} & \multicolumn{2}{c|}{$A_{u}$} &\multicolumn{3}{c|}{$B_{1g}$} & \multicolumn{3}{c|}{$B_{2g}$} & \multicolumn{3}{c}{$B_{3g}$} \\
     \multicolumn{1}{c}{LDA} & \multicolumn{1}{c}{PBEsol} & expt &  \multicolumn{1}{c}{LDA} & \multicolumn{1}{c}{PBEsol} &
         \multicolumn{1}{c}{LDA}& \multicolumn{1}{c}{PBEsol} & expt &   \multicolumn{1}{c}{LDA}& \multicolumn{1}{c}{PBEsol} & expt &
         \multicolumn{1}{c}{LDA}& \multicolumn{1}{c}{PBEsol} & expt  \\ \hline
      73.44 &  61.73 &  83  &  48.84 &  27.85 & 101.50 &  98.20  & 116  & 117.27 & 102.2  & 128   &   89.94 &   85.69 &  98 \\ 
     151.43 & 150.89 & 158  & 211.03 & 202.02 & 183.80 & 181.92  & 198  & 224.88 & 210.65 & 217   &  146.16 &  155.89 & 154 \\
     195.69 & 214.13 & 197  & 263.01 & 244.13 & 258.69 & 262.56  & 283  & 270.29 & 266.56 & 291   &  227.42 &  230.57 & 246  \\ 
     299.00 & 325.24 & 337  & 535.78 & 478.26 & 651.88 & 599.72  & 666  & 652.04 & 600.42 & 666   &  302.60 &  325.93 & 338 \\
     329.83 & 357.84 & 366  &        &        &        &         &      &        &        &       &  342.49 &  362.71 & 380 \\
     445.84 & 439.23 & 472  &        &        &        &         &      &        &        &       &  447.71 &  441.88 & 472 \\ 
     727.19 & 773.61 & 819  &        &        &        &         &      &        &        &       &  729.82 &  773.06 & 820  \\ 
     945.50 &1022.51 & 996  &        &        &        &         &      &        &        &       &  957.39 & 1028.89 & 996 \\ 
         \end{tabular}
  \end{ruledtabular}
  
\end{table*}

Our calculated values are compared with the experimental results of Seguin \etal\cite{Seguin1995} who also includes previous experimental results and provides a symmetry labeling
of the modes. However, we have relabeled them to take into account the
different choice of crystallographic axes here. Our $a,b,c$ correspond to Seguin's $c,a,b$.
Taking $x,y,z$ along $a,b,c$ this then also implies that
our $B_{1u},B_{2u},B_{3u}$
correspond to their $B_{2u},B_{3u},B_{1u}$
respectively and our $B_{1g},B_{2g},B_{3g}$
become their $B_{2g},B_{3g},B_{1g}$. $A_g$ and $A_u$ stay the same.
The $A_u$ modes are silent and can thus not be measured by either infrared or
Raman spectroscopies.

The calculated phonon frequencies are found to generally underestimate the
experimental ones with a few exceptions.
The largest absolute error in the LDA occurs for the $A_g^7$, $B_{3g}^7$  and the $B_{2u}^6$ modes, which are underestimated by about 80-100 cm$^{-1}$.
The error on these modes is somewhat reduced in PBEsol but is still of order 50 cm$^{-1}$. On the other hand, the PBEsol seems to underestimate
the low frequency modes significantly and its largest error now occurs for the $B_{3u}^3$, $B_{2g}^4$ and $B_{2g}^4$ modes.  The root mean square
error averaged over all TO modes and Raman modes is 39 cm$^{-1}$ in LDA and 29 cm$^{-1}$ in PBEsol, which is not a significant difference. 

Some modes have quite weak oscillator strengths and, where several modes are close in frequency, the experimental assignment
may not be entirely clear if polarization selection rules were not used.  For example for mode $B_{1u}^4$, the value 374 cm$^{-1}$ was
measured by Seguin \etal\cite{Seguin1995} while Py and Mashke \cite{Py81} give
a calculated value $B_{1uT}^4=380$ cm$^{-1}$ but did not observe it experimentally. 
Seguin \etal assign this mode as strong while
nearby $B_{2u}$ mode at 358 cm$^{-1}$ (363 cm$^{-1}$ in Py and Mashke\cite{Py81} as listed in Table \ref{tabmodes})is designated as weak. Another weak peak is observed in the IR spectrum at 350 cm$^{-1}$.
The oscillator strengths given in Table \ref{tabosc} show clearly
that $B_{1u}^4$ should be weaker than $B_{2u}^3$ and $B_{2u}^4$. The proximity of these modes makes it difficult to disentangle
them experimentally without using polarization dependence.

One may also observe that  each TO phonon mode of a given symmetry is
followed by an LO before the next TO phonon occurs. This is a general rule
obeyed by any crystal with at least orthorhombic symmetry,
but not for monoclinic symmetry. We note that this follows from general
considerations of the phonon related $\varepsilon$ and $\varepsilon^{-1}$
in a Lorentz oscillator model. 
\begin{figure}
  \includegraphics[width=4.1cm]{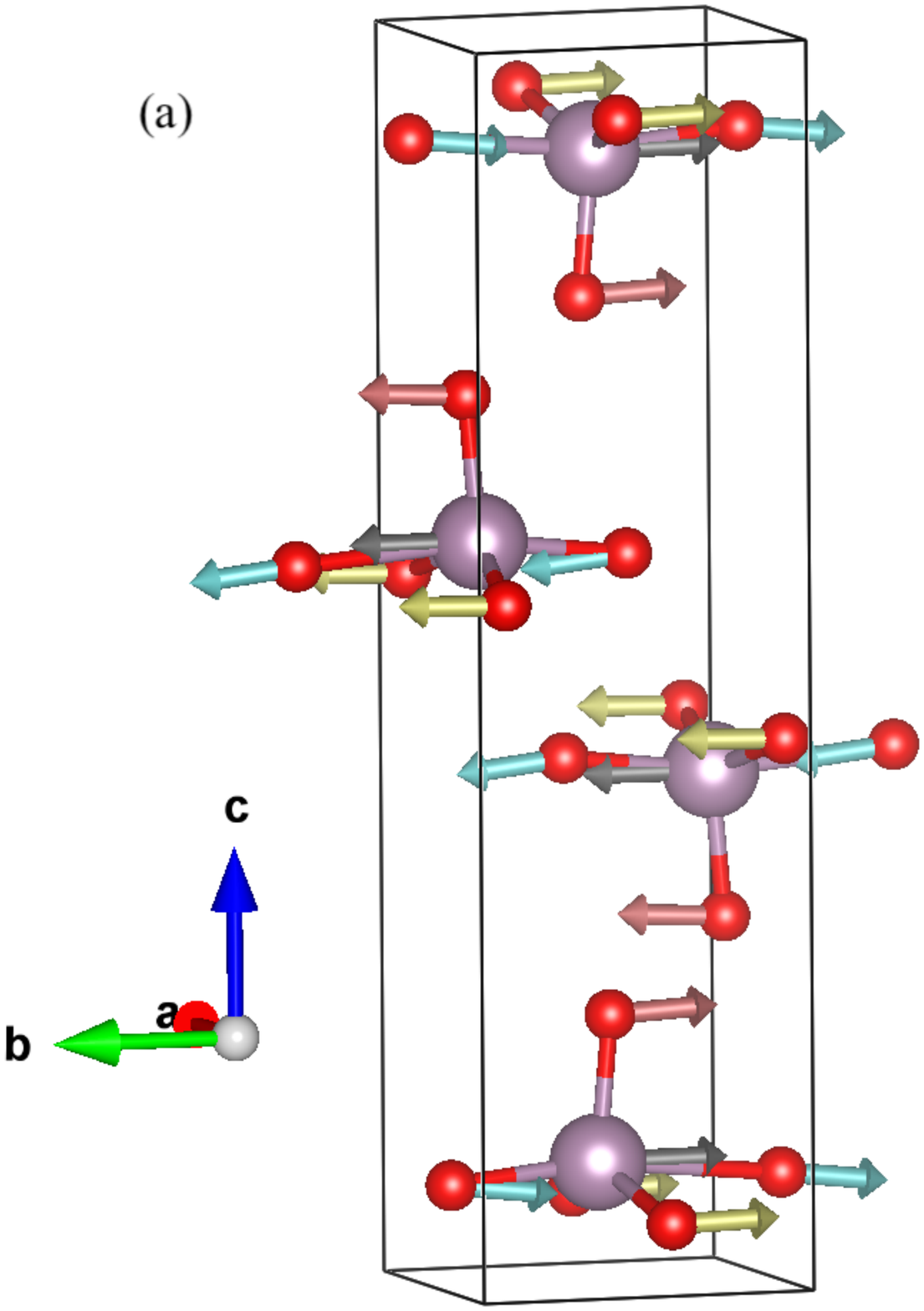}\includegraphics[width=3.9cm]{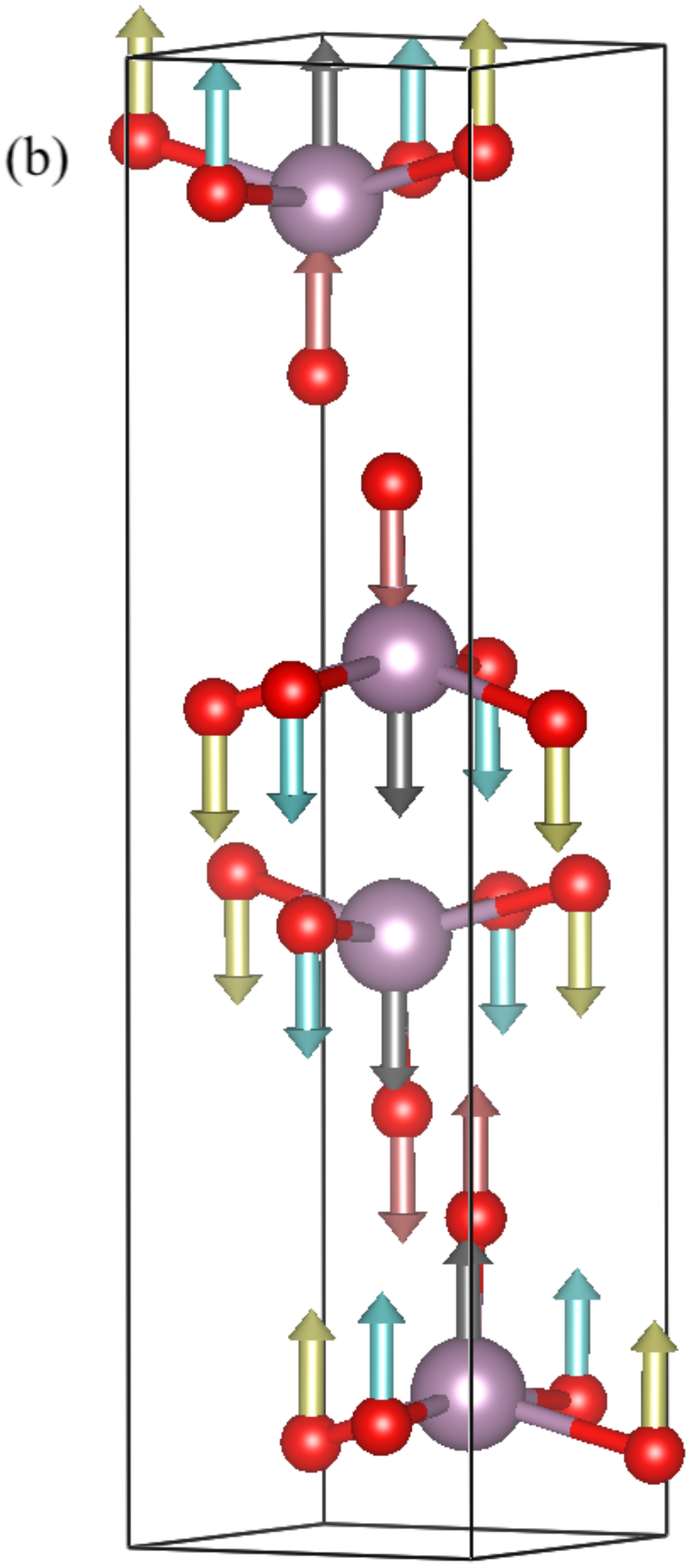}
  \caption{Eigendisplacement pattern for (a) $B_{1uT}^1$ mode and
    (b) $B_{2uT}^1$ modes. \label{fig:b1ut1b2ut1}}
\end{figure}

We now discuss the nature of a few of the  vibrational modes. 
The eigenvector displacements of all modes are given in Supplemental Information.\cite{SM}
The lowest frequency $B_{1u}^1$ corresponds to a sliding of an entire bilayer with respect to the other in the $b$ direction as can be seen in
Fig. \ref{fig:b1ut1b2ut1}(a).
The $B_{2u}^1$ mode on the other hand has bilayers moving relative to each other perpendicular to each other Fig. \ref{fig:b1ut1b2ut1}(b).
One may expect these modes to be rather sensitive to the weak van der Waals like interlayer coupling. Because in PBEsol,
the layers are somewhat farther apart, these mode frequencies are underestimated.

\begin{figure}
  \includegraphics[width=4cm]{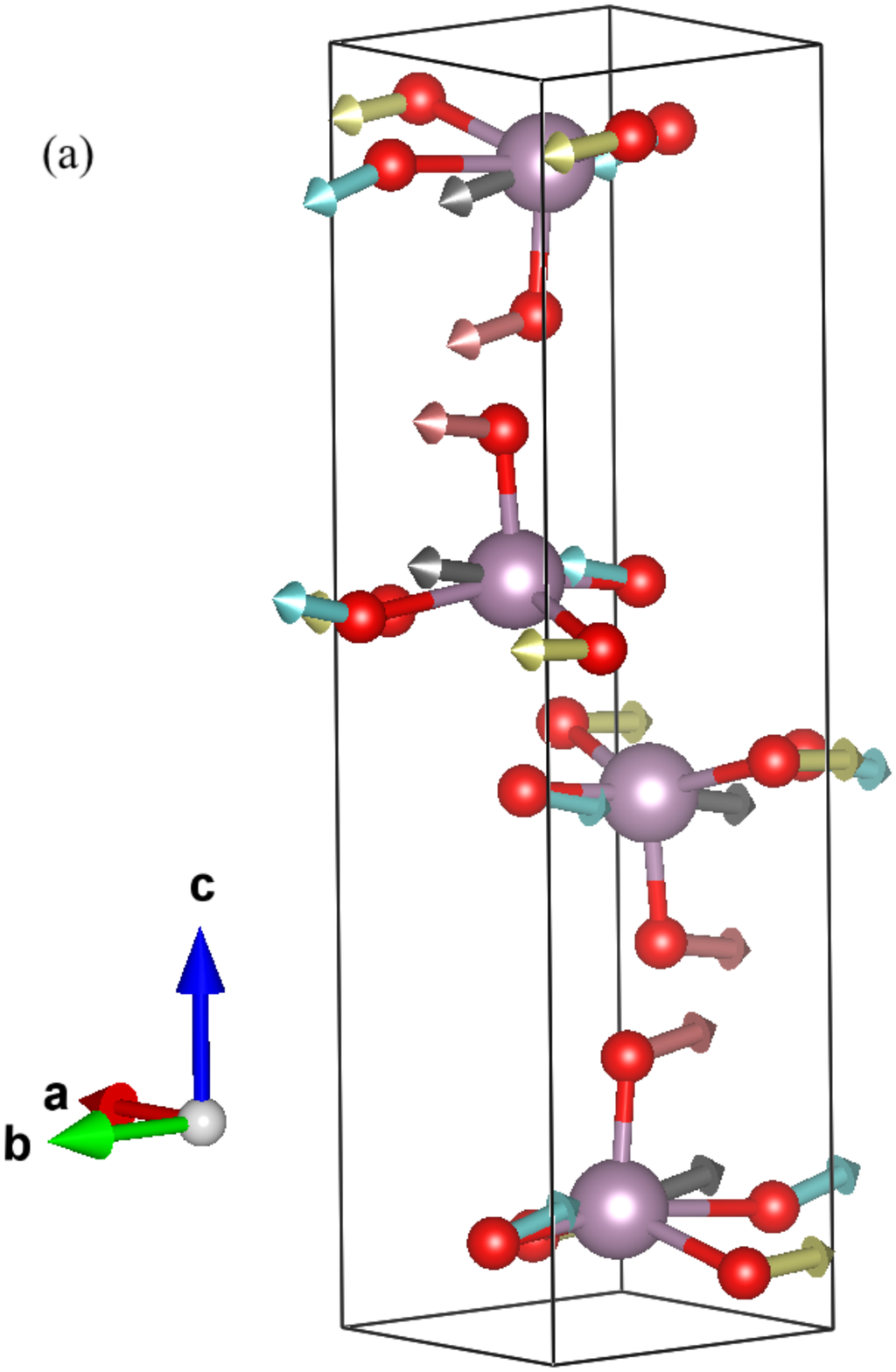}\includegraphics[width=4cm]{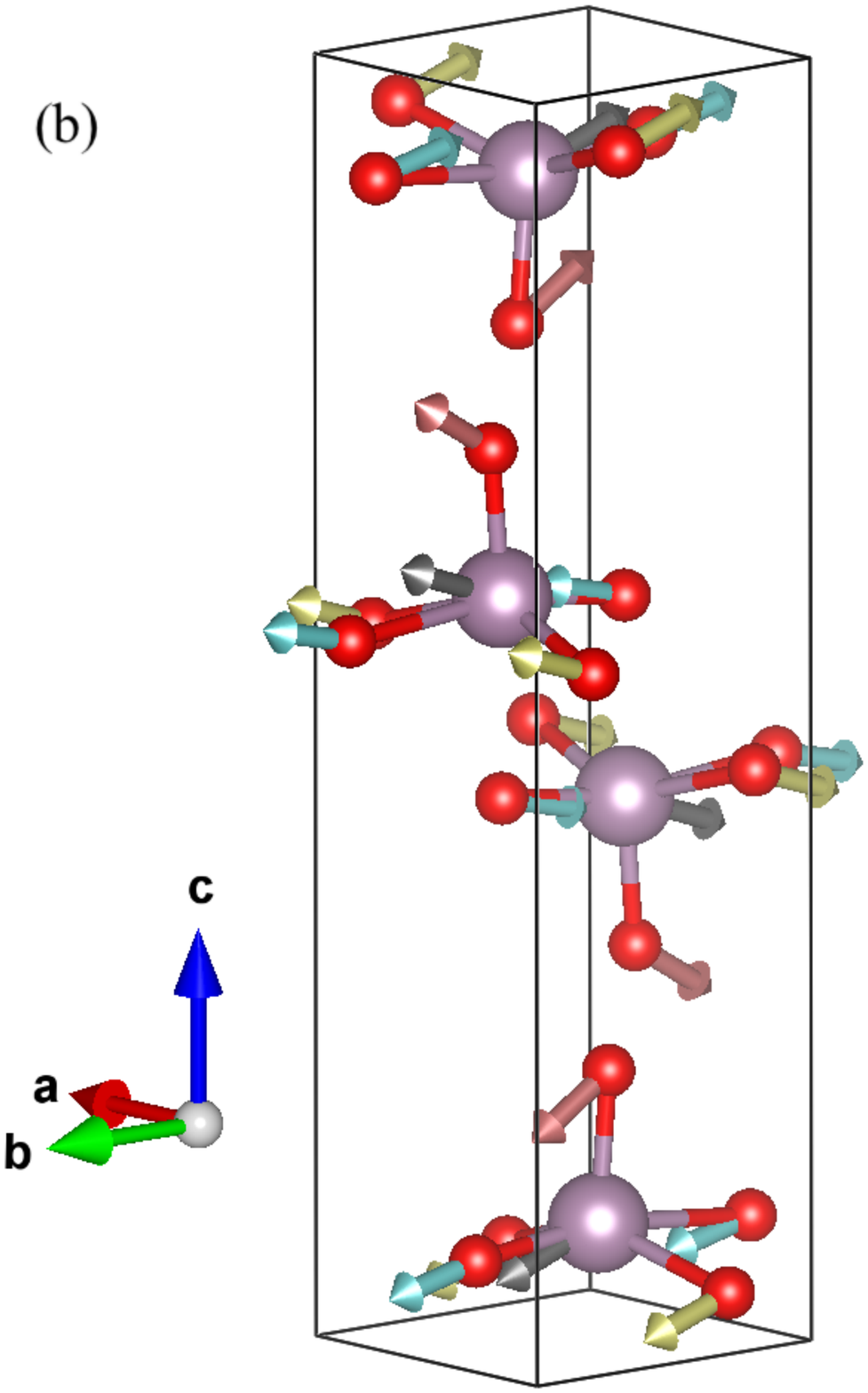}
  \caption{Eigendisplacement pattern for (a)  $A_g^1$ mode
    and (b) $B_{3g}^1$ mode.\label{fig:ag1}}
\end{figure}
The $A_g^1$ mode on the other hand consists mostly of a sliding of the layers within one bilayer with respect to each other
but also with a slight breathing component of the distance between these layers within the bilayer (Fig. \ref{fig:ag1})(a).  This mode
is already significantly higher in frequency which clearly shows that the bonding between layers within a bilayer
is stronger than between bilayers. 
The lowest $B_{3g}^1$ mode is similar but with the two bilayers having opposite sign instead of the same sign (Fig. \ref{fig:ag1}(b)).
The $A_g^2$ and $B_{3g}^2$ modes
are mostly a breathing mode of the interlayer distance within a bilayer but again, either in phase between the two bilayers or out of phase.
The intermediate frequency modes are more complex in nature.

\begin{figure}
  \includegraphics[width=4cm]{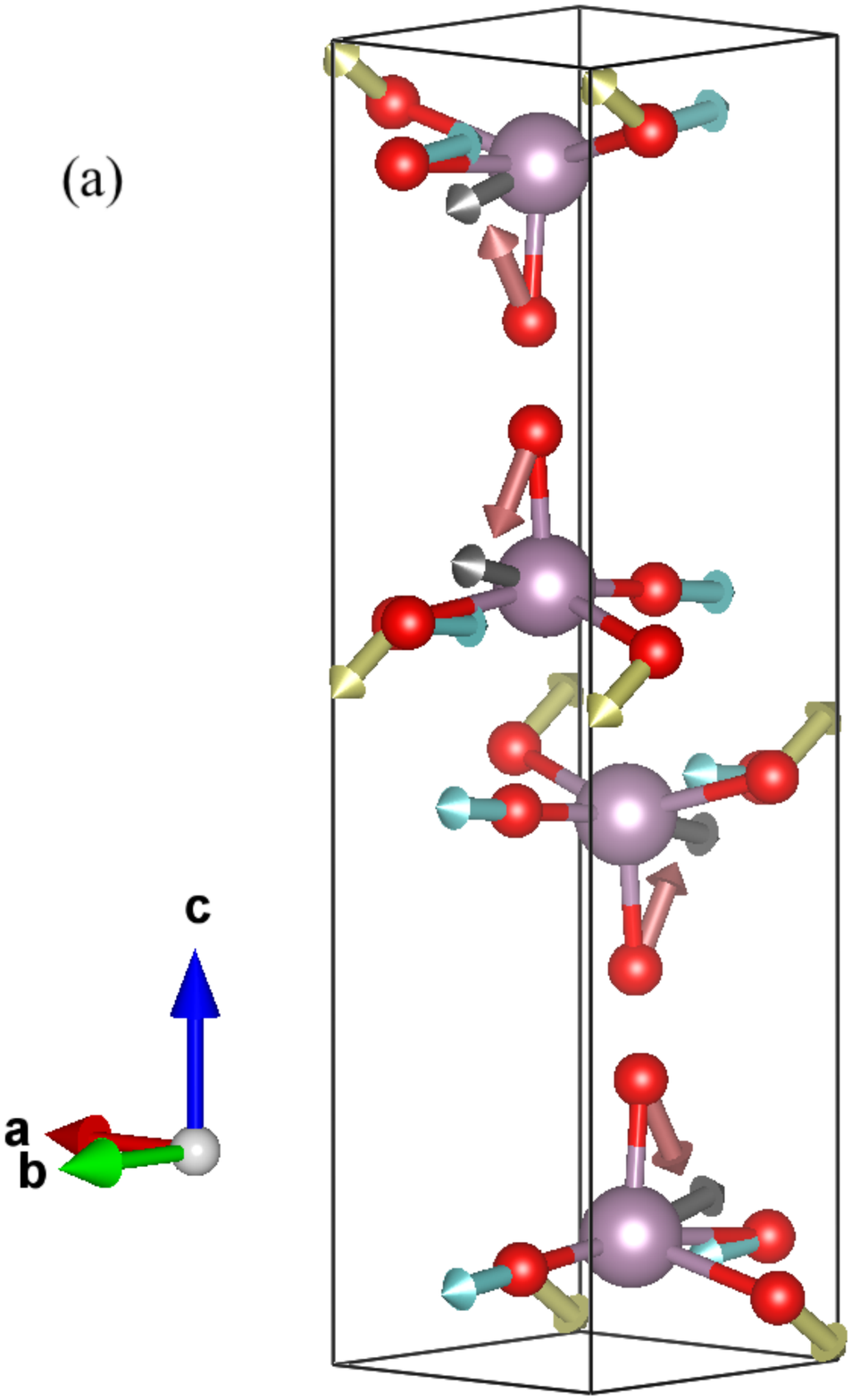} \includegraphics[width=4cm]{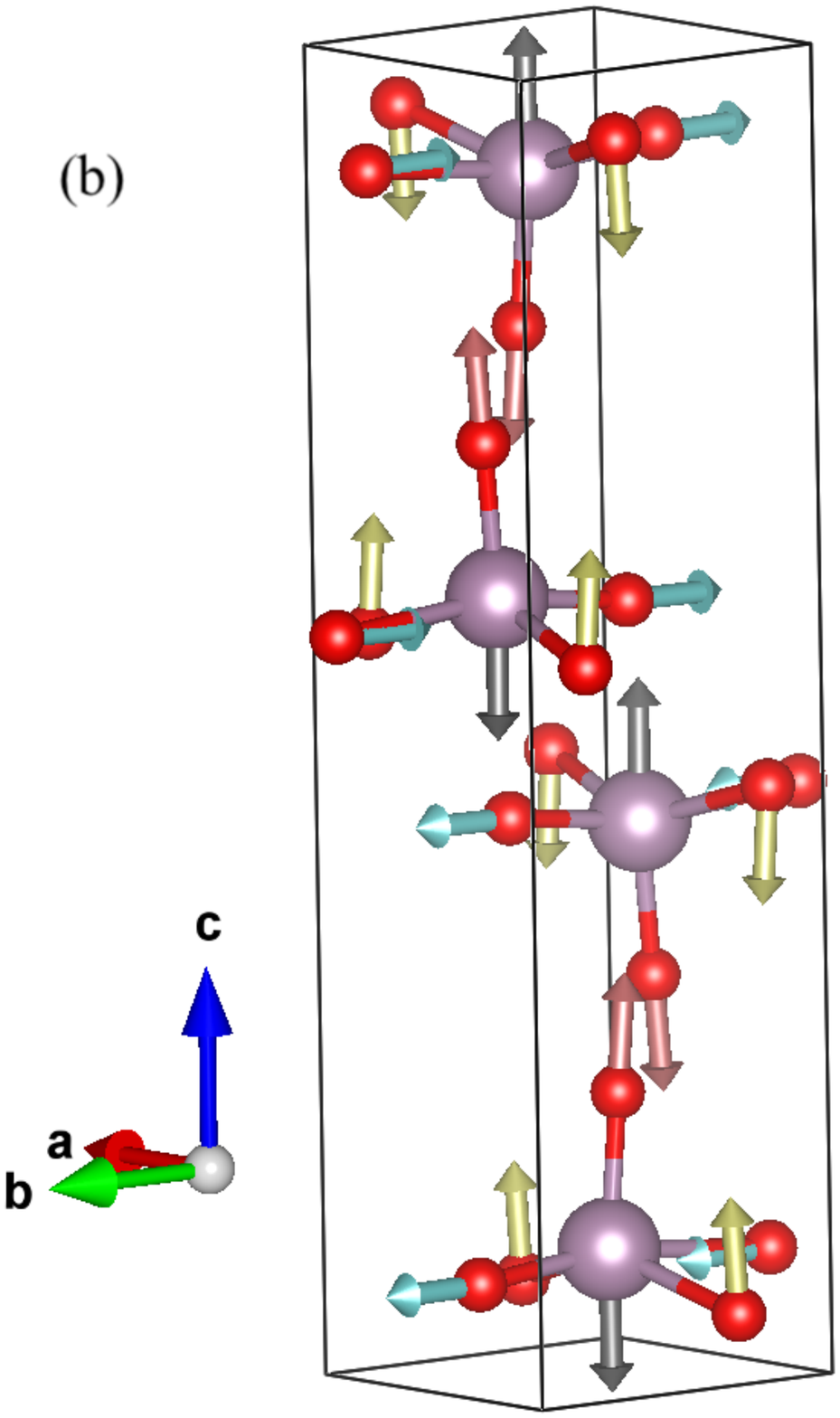}
  \caption{Eigendisplacement pattern for (a)  $A_g^7$ mode
    and (b) $A_g^8$ mode.\label{fig:ag78}}
\end{figure}
However, the $A_g^7$ mode shows a strong Mo-O$^{(1)}$ bond stretch
character with also some Mo-O$^{(2)}$ stretch character, while $A_g^8$ is characterized by a stretch of the short Mo-O$^{(2)}$ bond
which explains why this mode has one of the highest frequencies.
(See Figs. \ref{fig:ag78}(a) and (b).)

It may be noticed that several  modes are grouped in groups of four modes
with frequencies close to each other. This is because the same local pattern can either be in phase or out of phase
between the two Mo within a bilayer and between the two bilayers.  Thus for example there are four modes close to 814 cm$^{-1}$
using the experimental value, they are  the $A_g^7$, $B_{3g}^7$,  $B_{1u}^6$, $B_{2u}^6$ modes. All of these are significantly
underestimated and occur near 730 cm$^{-1}$ in LDA and near 770 cm$^{-1}$ in PBEsol. Because these modes involve strong motion
along the $y=b$ direction, it has a strong coupling to an electric field along $y$ for the $B_{2uL}$ mode which occurs at 974 cm$^{-1}$.
Similarly, there are four high frequency modes near 1000 cm$^{-1}$.

We find that the modes near 730 cm$^{-1}$, ($A_g^7$, $B_{3g}^7$,  $B_{1u}^6$, $B_{2u}^6$) are quite sensitive to the
interlayer distance. We can see this by comparing the PBEsol results at the PBEsol lattice constants  with
the PBEsol results at the PBE lattice constants which have respectively a $c$ lattice constant of 16.919 \AA\  and 14.425 \AA.
We find the phonon frequency decreases by $\sim$40 cm$^{-1}$ by using the larger interlayer distance. This suggests
that further decreasing the lattice constant closer to experiment would reduce the error in this mode frequency. 
On the other hand, the highest modes increase only slightly
in frequency  (by about 4 cm$^{-1}$)
when using the larger interplanar distance.  This may also reduce the apparent overestimate of this mode by the PBEsol calculation.


\subsection{Infared spectra and associated quantities.}\label{infrared}
\begin{figure}
 \includegraphics[width=8cm]{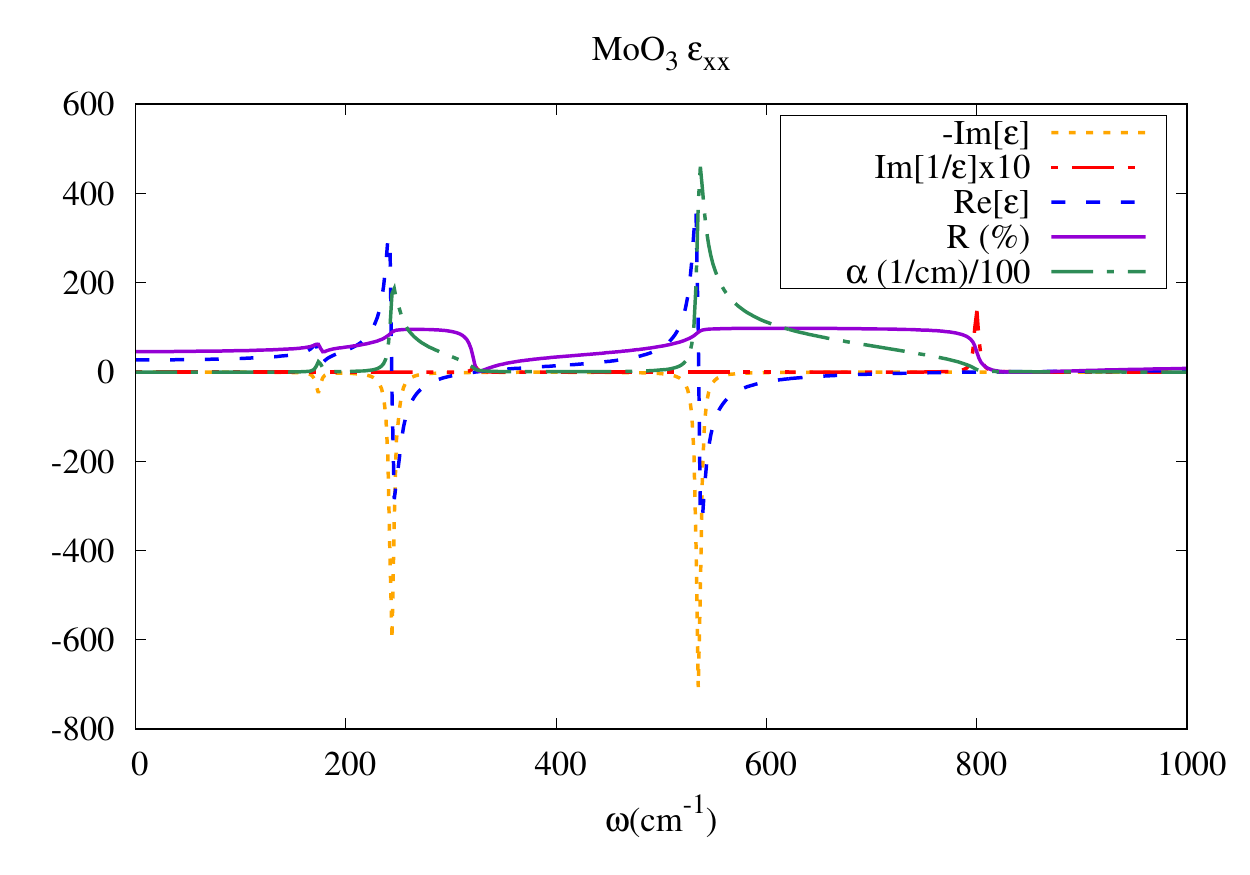}
 \caption{IR spectra for $B_{3u}$ modes.\label{figirx}}
\end{figure}

\begin{figure}
 \includegraphics[width=8cm]{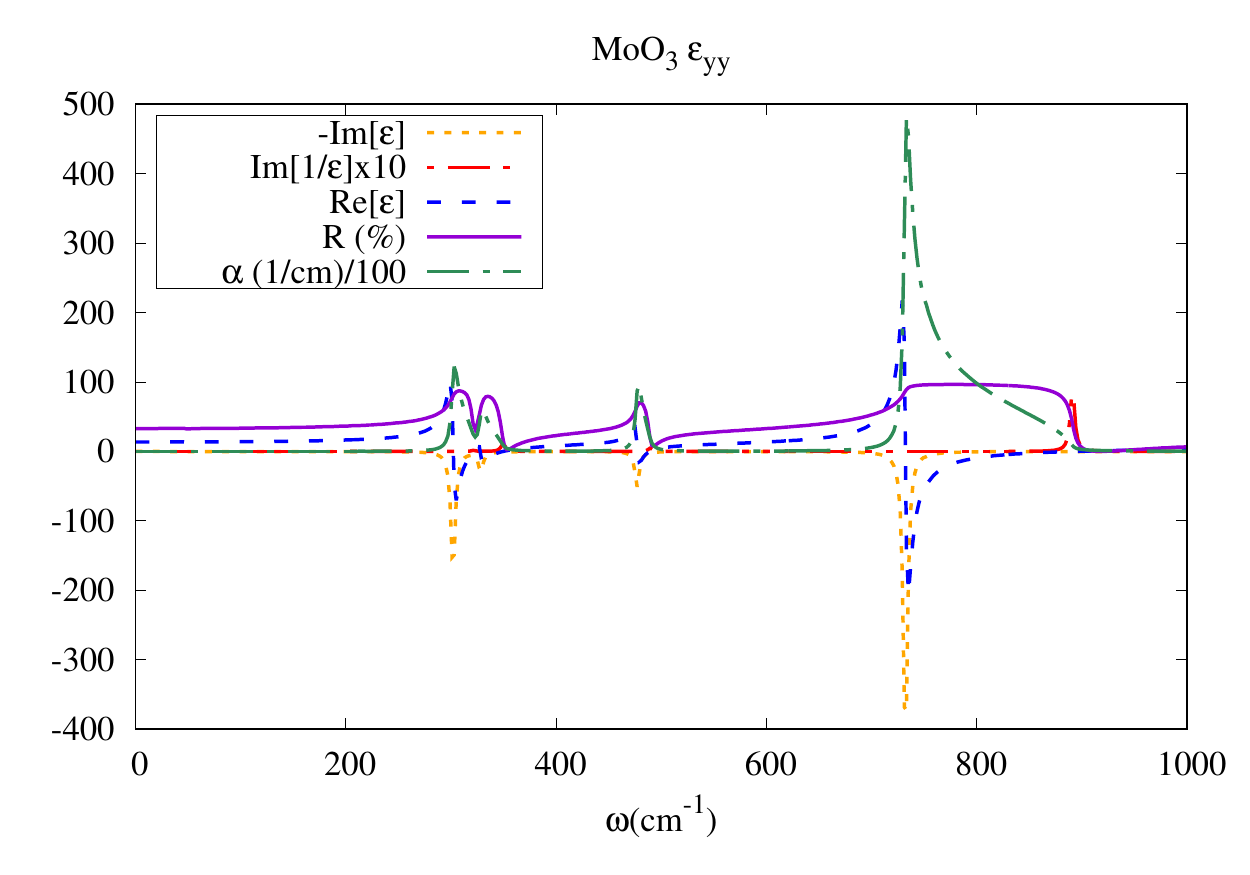}
 \caption{IR spectra for $B_{2u}$ modes.\label{figiry}}
\end{figure}

\begin{figure}
 \includegraphics[width=8cm]{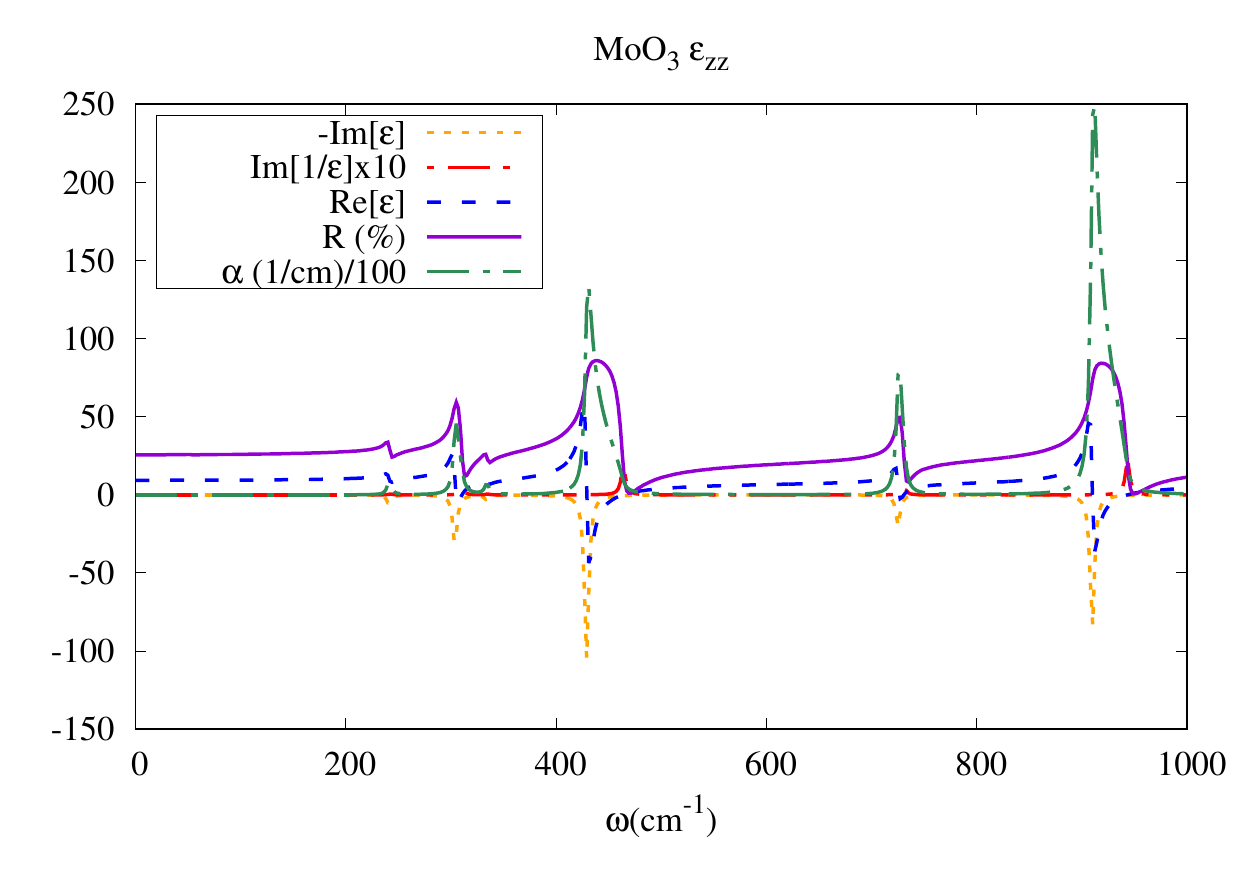}
 \caption{IR spectra for $B_{1u}$ modes.\label{figirz}}
\end{figure}

\begin{table}
  \caption{Born effective charge tensor elements
    for the atoms listed in Table \ref{tabstruc}.
    By symmetry all $Z^*_{xy}$, $Z^*_{yx}$, $Z^*_{xz}$, $Z^*_{zx}$ are zero. \label{tabBorn}}
  \begin{ruledtabular}
  \begin{tabular}{ccccccc} \\  
     Components(label) & Mo & O$^{(1)}$ & O$^{(2)}$ & O$^{(3)}$  \\
     \hline
     \multicolumn{5}{c}{LDA} \\ \hline 
     $Z^{*}_{xx}(A_{g})$ & 7.483 & $-1.139$  & $-0.552$ & $-5.790$ \\
     $Z^{*}_{yy}(A_{g})$ &  6.649 & $-4.859$  & $-0.619$ &$-1.169$ \\
     $Z^{*}_{zz}(A_{g})$ & 4.571  & $-0.686$ & $-2.275$ & $-1.609$ \\
     $Z^{*}_{yz}(B_{3g})$ &  0.285 &  $-0.305$ & $-0.343$ & $-0.303$ \\
     $Z^{*}_{zy}(B_{3g})$ & 0.617 & $-0.506$ & $-0.350$ & $-0.197$  \\ \hline
     \multicolumn{5}{c}{PBEsol} \\ \hline
      $Z^{*}_{xx}(A_{g})$ & 7.632 & $-1.115$ & $-0.580$  & $-5.940$ \\
     $Z^{*}_{yy}(A_{g})$ &  6.302 & $-4.600$ & $-0.601$ &$-1.104$ \\
     $Z^{*}_{zz}(A_{g})$ & 4.268  & $-0.563$ & $-2.172$ & $-1.532$ \\
     $Z^{*}_{yz}(B_{3g})$ & 0.290 & $-0.304$ & $-0.377$ & $-0.354$ \\
     $Z^{*}_{zy}(B_{3g})$ & 0.553  & $-0.438 $ & $-0.361$ & $-0.223$  \\ \hline
            \end{tabular}
  \end{ruledtabular}
\end{table}

In this section, we present our simulated infrared spectra and
associated quantities. All data reported here were obtained from the LDA calculation. 
These are obtained from calculating the contribution of phonons
to the dielectric response function in terms of the classical Lorentz
oscillator model. Within DFPT, the oscillator strengths can be obtained
directly from the phonon eigenvectors and the Born effective charges,
which describe the coupling of the vibrational modes to an
electric field and are obtained as a mixed derivative of the total
energy {\sl vs}. a static electric field and an atomic displacement, given by 
\begin{equation}
  Z^*_{\kappa\beta\alpha}=\frac{\partial^2 E_{tot}}{\partial u_{\kappa\alpha}\partial E_\beta}=V\frac{\partial P_\beta}{\partial u_{\kappa\alpha}}=\frac{\partial F_{\kappa\alpha}}{\partial E_\beta}
\end{equation}
where $P_\beta$ is the macroscopic polarization, $V$ the unit cell volume
and $u_{\kappa\alpha}$ the displacement of atom $\kappa$ in direction $\alpha$
which for a ${\bf q}=0$ mode  is the same in each unit cell. $F_{\kappa\alpha}$
is the force on the atom $\kappa$ in direction $\alpha$ and $E_\beta$
is the electric field component. Atomic units are used throughout in which
$\hbar=e=m_e=1$.
Note that the Born effective charge tensors are not macroscopic tensors
but only reflect the point group symmetry of the Wyckoff site of that
atom. Because the atoms are all in $4c$ positions which lie on
the mirror planes $m_x$ and hence need to have zero $xz$ and $xy$ tensor
elements. However they do have a non-zero $yz$ and $zy$ element, which
differ because the first index refers to the derivative {\sl vs.} electric
field and the second to the derivative {\sl vs.} atom displacement direction.
The Born charges are seen to deviate significantly from the nominal
charge of Mo$^{+6}$ and O$^{-2}$ and have also significant anisotropies.
Specifically, O$^{(2)} $ which is bonded to a single Mo in the $z$ direction is seen
to be anomalously small in the $x$ and $y$ directions. On the other hand O$^{(1)}$
which is the bridge oxygen is seen to have the largest effective charge component  in the $y$
direction and O$^{(3)}$ in the $x$ direction. 
The off-diagonal $yz,zy$ elements sum to zero for each atom type separately
because of the sign changes of the symmetry related atoms  which behave
as $B_{3g}$. The diagonal terms sum to zero for each diagonal component
when summing over all atoms, balancing the cation and anions.

The oscillator strength is then given by 
\begin{equation}
S_{n,\alpha\alpha} = \left| \sum_{\kappa,\alpha'} Z^{*}_{\kappa,\alpha\alpha'}U_{n}(\kappa,\alpha') \right|^{2},
\end{equation}
where $Z^{*}_{\kappa,\alpha\alpha'}$ are the Born effective
charge tensor components given in Table \ref{tabBorn}, $U_{n}(\kappa,\alpha')$ are the eigenvectors for each of the modes $n$ at ${\bf q}=0$
and, $\kappa$ refers to the atom label. The eigenvectors are normalized
as 
\begin{equation}
  \sum_{\kappa\alpha} U_n(\kappa,\alpha)^*M_\kappa U_m(\kappa,\alpha)=\delta_{nm},
\end{equation}
where $M_\kappa$ are the atom masses. 
Note that because of the orthorhombic
symmetry the oscillator strength tensor is diagonal. 
Its non-zero elements are listed in Table \ref{tabosc}. 
One can see from this table, that the higher frequency modes tend to
have higher oscillator strengths. This is  because they correspond
to bond stretches and thus have a significant dipole moment associated
with them. An exception is the highest $B_{2u}^7$ mode has quite small oscillator
strength and correspondingly also small TO-LO splitting. 

\begin{table}
  \caption{Oscillator strength tensors.\label{tabosc}}
  \begin{ruledtabular}
  \begin{tabular}{ccc} \\ 
      $S_{n,xx}$ $(B_{3u})$ & $S_{n,yy}$ $(B_{2u})$ & $S_{n,zz}$ $(B_{1u})$  \\
     \hline
     $9.07 \times 10^{-5}$ & $3.05 \times 10^{-7}$ & $6.94 \times 10^{-8}$ \\
     $1.67 \times 10^{-3}$ & $ 4.85\times 10^{-7}$ & $1.559 \times 10^{-5}$ \\
     $4.33 \times 10^{-3}$ & $5.94 \times 10^{-4}$ & $1.08 \times 10^{-4}$ \\
                                        & $1.06 \times 10^{-4}$ & $1.07 \times 10^{-5}$ \\
                                        & $2.73 \times 10^{-4}$ & $5.05 \times 10^{-4}$ \\
                                        & $3.55 \times 10^{-3}$ & $1.62 \times 10^{-4}$ \\
                                        & $2.22 \times 10^{-9}$ & $8.71 \times 10^{-4}$ \\
     
       \end{tabular}
  \end{ruledtabular}
\end{table}

The frequency dependent dielectric function in the region below the
band gap is given by 
\begin{equation}
\varepsilon_{\alpha\alpha}(\omega)=\varepsilon_{\alpha\alpha}^{\infty}+\frac{4 \pi}{V}\sum_{n}\frac{S_{n,\alpha\alpha}}{\omega_n^2-\omega^2-i\Gamma_n\omega}
\end{equation}
where $\omega_n$ are the phonon frequencies and $\Gamma_n$ is a damping factor. The latter is not calculated and we just
assign a uniform value of 5 cm$^{-1}$ to it for all modes. 

The first term $\varepsilon^\infty$  is the high-frequency dielectric constant, meaning
at frequencies below the gap but above the phonon frequencies. 
More precisely it is the static limit of the electronic contribution to the
dielectric function, in other words the contribution from all higher frequency
excitations, namely the inter-band optical transitions.
It is calculated in the DFPT framework as the adiabatic response
to a static electric field in the $x$, $y$, $z$
directions. Because of the orthorhombic symmetry it is also a diagonal tensor, 
$\varepsilon_{\alpha\alpha}^\infty$. The values of this tensor are given in
Table \ref{tabepsinf}. 
They  are directly related to the anisotropic indices of refraction 
in the visible region below
the gap but above the phonon frequencies. The values of 
$n_{\alpha\alpha}=\sqrt{\varepsilon_{\alpha\alpha}^\infty}$ are given
in Table \ref{tabindex} for convenience.
The static dielectric constant $\varepsilon^0_{\alpha\alpha}$ in Table \ref{tabepsinf}
applies for frequencies well below
the phonon frequencies.

\begin{table}
  \caption{High-frequency and static dielectric tensor components.\label{tabepsinf}}
  \begin{ruledtabular}
  \begin{tabular}{lcccccc} \\  
  method &   $\varepsilon^{\infty}_{xx}$ & $\varepsilon^{\infty}_{yy}$ & $\varepsilon^{\infty}_{zz}$ & $\varepsilon^{0}_{xx}$ & $\varepsilon^{0}_{yy}$ & $\varepsilon^{0}_{zz}$ \\
     \hline
LDA    &      6.792 & 6.162 & 4.662 & 27.210 & 13.024 & 7.173  \\
PBEsol &     5.959 & 5.205 & 4.001 &        &        &        \\
   \end{tabular}
  \end{ruledtabular}
\end{table}

\begin{table}
  \caption{The indices of refraction.\label{tabindex}}
  \begin{ruledtabular}
  \begin{tabular}{lccc} \\  
    method & $n_{xx}$ & $n_{yy}$ & $n_{zz}$ \\
     \hline
   LDA &   2.606 & 2.482 & 2.159 \\
   PBEsol &   2.441 & 2.282 & 2.000 \\
  \end{tabular}
  \end{ruledtabular}
\end{table}

\begin{figure}
  \includegraphics[width=8cm]{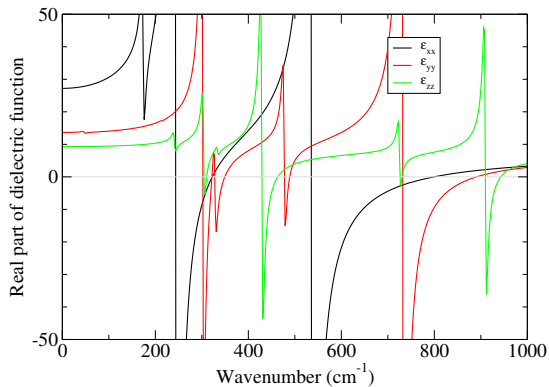}
\caption{Real part of dielectric function for different directions.\label{figepsxyz}}
  \end{figure}

From the above defined $\varepsilon(\omega)$ we can extract
various related optical functions, in the infrared range. 
In particular, the optical absorption 
 $\alpha(\omega)=2\omega\mathrm{Im}[\varepsilon(\omega)]/n(\omega)$ and
the  reflectivity $R(\omega)=|(\tilde n(\omega)-1)/(\tilde n(\omega)+1)|^2$
with $\tilde n=n+i\kappa=\sqrt{\varepsilon}$ the complex index of refraction
as well as the loss function $-\mathrm{Im}\{\varepsilon(\omega)^{-1}\}$
are the most closely related to the measurements. 
The zeros in the real part $\varepsilon_1(\omega)=\mathrm{Re}[\varepsilon(\omega)]$ and the peaks
in  the loss function indicate the LO mode frequencies, while the peaks
in $\varepsilon_2(\omega)=\mathrm{Im}\{\varepsilon(\omega\}$ give the TO modes.
The reflectivity shows the typical {\sl Reststrahlen} bands (RB)
which jump to almost 100\% reflectivity at the TO modes and
fall back at the LO modes. 
Note that the absorption coefficient  shows peaks corresponding to those
in $\varepsilon_2(\omega)$ but also shoulders at the zeros of $\varepsilon_1(\omega)$.
The infrared spectra for the three polarizations are shown in
Figs. 
\ref{figirx},\ref{figiry},\ref{figirz}. 
These correspond
respectively to $B_{1u}$, $B_{2u}$ and $B_{3u}$ modes which are active
for polarizations along $z$, $y$ and $x$.

We may compare these with the IR absorption spectra of
Seguin \etal\cite{Seguin1995} which however do not mention the polarization.
The highest absorption band found by them near 1000 cm$^{-1}$
agrees well with our $B_{1u}$ peak at 909-945 cm$^{-1}$ and corresponds to
$z$-polarization, related to the Mo-O$^{(2)}$ bond stretch of the shortest bond.
The next main feature in Seguin \etal\cite{Seguin1995} corresponds to our
$B_{2u}$ spectrum for $y$-polarization and starts at $B_{2uT}^6$ at 732 cm$^{-1}$ 
and ends at the $B_{2uL}^6$ at 906 cm$^{-1}$. Note that this mode is also
close to the  strongest  $A_g$ mode in Raman.  
However, the sharp feature on that peak at lower energy 
with much smaller LO-TO splitting is the $B_{1uT}^6,B_{1uL}^6$ RB.
The next broad feature is clearly dominated by the $B_{3u}^3$ RB
between 535 cm$^{-1}$ and 799 cm$^{-1}$.  In the lower frequency region, a RB
occurs near 260 cm$^{-1}$ in the experiment, which corresponds to peaks in our spectra
near 240 cm$^{-1}$ and stems mostly from the $x$ polarization $B_{3u}^2$ mode.
A less intense RB is seen near 350 cm$^{-1}$ which corresponds to our $B_{2u}^4$ mode.
\begin{table}
  \caption{Frequency ranges (cm$^{-1}$)  with real $\varepsilon$ of different
    sign in different direcions, signs as indicated.\label{tabepsxyz}}
  \begin{ruledtabular}
    \begin{tabular}{llccc}
      $\omega_{min}$ & $\omega_{max}$ & $\varepsilon_{xx}$ & $\varepsilon_{yy}$ & $\varepsilon_{zz}$ \\ \hline
      243 & 302 & $-$ & $+$ & $+$ \\
      310 & 322 & $-$ & $-$ & $+$ \\
      327 & 350 & $+$ & $-$ & $+$ \\
      428 & 462 & $+$ & $+$ & $-$ \\
      477 & 488 & $+$ & $-$ & $+$ \\
      537 & 727 & $-$ & $+$ & $+$ \\
      733 & 794 & $-$ & $-$ & $+$ \\
      907 & 943 & $+$ & $+$ & $-$ \\
    \end{tabular}
  \end{ruledtabular}
\end{table}

The anisotropy is important for this material. It leads to various
ranges of wavenumber where $\varepsilon_1(\omega)$ has negative sign
in one or two directions and positive in the other direction(s). This implies
the material is hyperbolic in its dispersion in these ranges.
Combined with low losses in these regions, or small imaginary part, this
allows for interesting optical applications based on phonon-polaritons in
the mid-infrared range.\cite{Ma2018,Dixit2021} Fig. \ref{figepsxyz}
show the real part of the dielectric functions for the three directions
together. One can see various ranges where the dielectric constant
has opposite sign in different directions.  These are summarized in Table \ref{tabepsxyz}.
We should caution that we here have used an arbitrary broadening
factor $\Gamma_n$
in the calculation of the dielectric function. Therefore we cannot
at present accurately estimate the width of the peaks in the imaginary
part $\varepsilon_2(\omega)$ which, in this context, is important
in gauging the losses in propagating light.

\subsection{Raman spectra}\label{ramanspec}
\begin{table}
  \caption{Raman tensor components.\label{tabraman}}
  \begin{ruledtabular}
  \begin{tabular}{ccc} 
     $A_{g}$ & $A_{g}$ & $A_{g}$ \\
      $a=\alpha_{xx}$ & $b=\alpha_{yy}$ &  $c=\alpha_{zz}$  \\
     \hline
     $1.99 \times10^{-3}$ &  $3.34  \times 10^{-3}$ & $-6.09 \times 10^{-5}$ \\
     $-6.93 \times10^{-3}$ &  $-3.85  \times 10^{-3}$ & $-2.39 \times 10^{-4}$ \\
     $3.84 \times10^{-3}$ &  $1.83  \times 10^{-3}$ & $-2.69 \times 10^{-3}$ \\
     $1.04 \times10^{-2}$ &  $1.16  \times 10^{-2}$ & $-1.22 \times 10^{-4}$ \\
     $-6.95 \times10^{-3}$ &  $3.32  \times 10^{-3}$ & $-3.98 \times 10^{-4}$ \\
     $2.97 \times10^{-3}$ &  $4.26  \times 10^{-3}$ & $-9.03 \times 10^{-3}$ \\
     $3.20 \times10^{-2}$ &  $7.01  \times 10^{-2}$ & $4.84 \times 10^{-4}$ \\    
     $1.52  \times 10^{-2}$ & $1.90  \times 10^{-2}$ & $3.04 \times 10^{-2}$ \\\hline
     
        $B_{1g}$          &    $B_{2g}$         &      $B_{3g}$ \\
      $d=\alpha_{xy}$ & $e=\alpha_{xz}$ & $f=\alpha_{yz}$  \\
     \hline
     $3.24 \times10^{-3}$ &  $ -3.09 \times 10^{-3}$ & $1.00 \times 10^{-3}$ \\
     $-3.92 \times10^{-3}$ &  $3.67  \times 10^{-3}$ & $7.40 \times 10^{-4}$ \\
     $1.17 \times10^{-2}$ &  $-9.62  \times 10^{-3}$ & $1.33 \times 10^{-3}$ \\
     $9.17 \times10^{-3}$ &  $-1.72  \times 10^{-2}$ & $2.16 \times 10^{-4}$ \\
                                       &                                       &  $-7.88 \times 10^{-3}$ \\
                                       &                                       &  $1.26 \times 10^{-3}$ \\
                                        &                                      &  $8.75 \times 10^{-4}$ \\
                                       &                                       &  $-9.53 \times 10^{-4}$ \\ \hline
       \end{tabular}
  \end{ruledtabular}
\end{table}

The Raman cross-section  for the Stokes process (energy loss)
for each mode is given by,
\begin{equation}
  \frac{dS}{d\Omega}=\frac{(\omega_0-\omega_m)^4}{c^4}|{\bf e}_i\cdot\bm{\alpha}^m\cdot{\bf e}_o|^2\frac{\hbar}{2\omega_m}(n_m+1)
\end{equation}
where $\omega_0$ is the incident light frequency, $\omega_m$
the mode frequency, and $n_m$ is the
phonon occupation number $n_m=[\exp{(\hbar\omega_m/k_BT)}-1]^{-1}$,
${\bf e}_{i}$ and ${\bf e}_{o}$ refer to the incident and the
scattered polarization directions and $\bm{\alpha}^m$ is the second-rank Raman susceptibility tensor for mode $m$ which is given by,
\begin{equation}
\alpha^{m}_{\alpha\beta} = \sqrt{V}\sum_{\kappa\gamma}\frac{\partial \chi_{\alpha\beta}}{\partial \tau_{\kappa \gamma}} U_{m}(\kappa \gamma),
\end{equation}
in terms of  $U_{m}(\kappa \gamma)$,the eigenvector of the $m$-th vibrational mode and the derivative of the susceptibility {\sl vs.} atomic displacements. 

\begin{figure}
 \includegraphics[width=8cm]{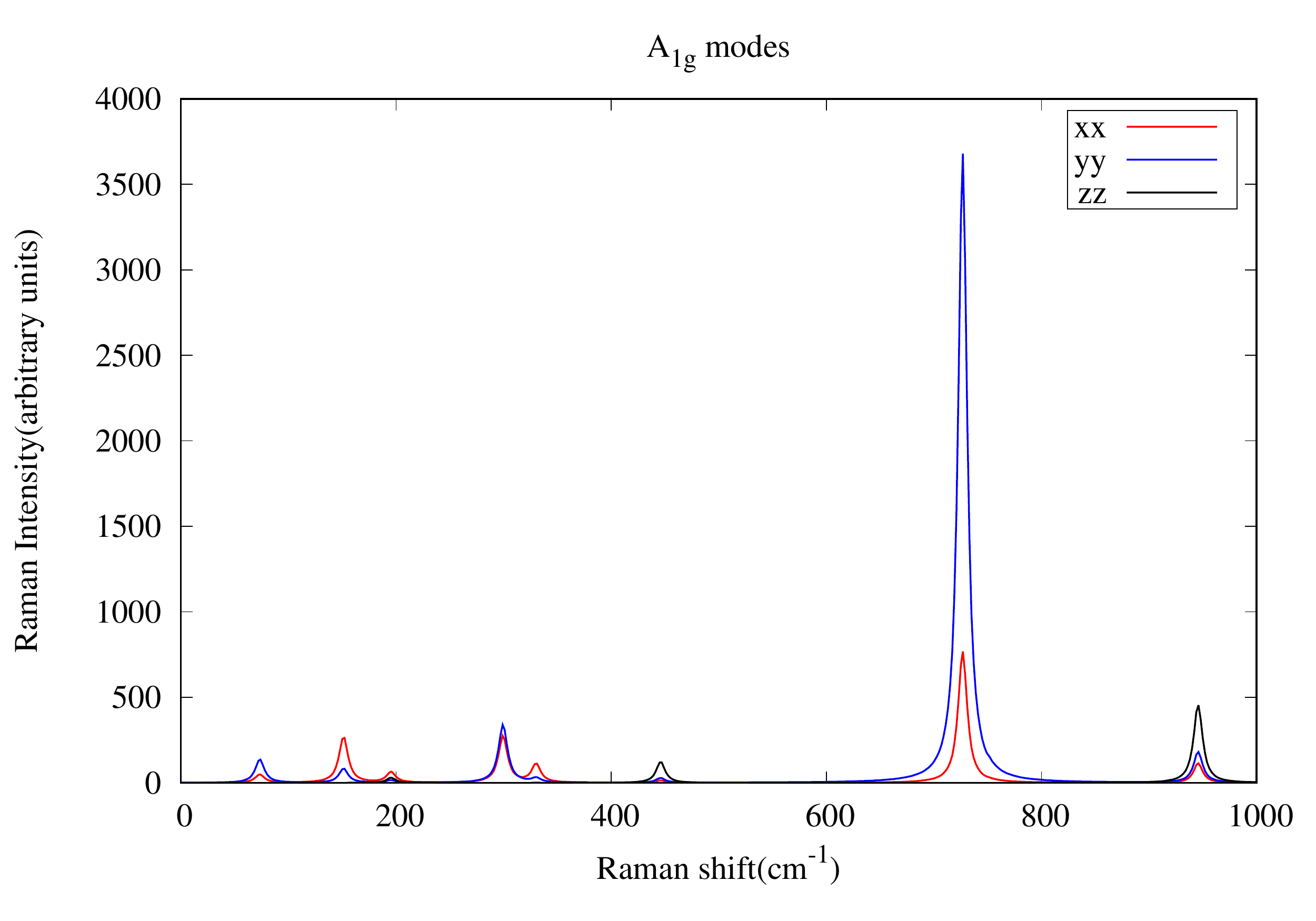}
 \caption{Raman spectra for $A_{g}$ modes.\label{figramanag}}
\end{figure}

\begin{figure}
 \includegraphics[width=8cm]{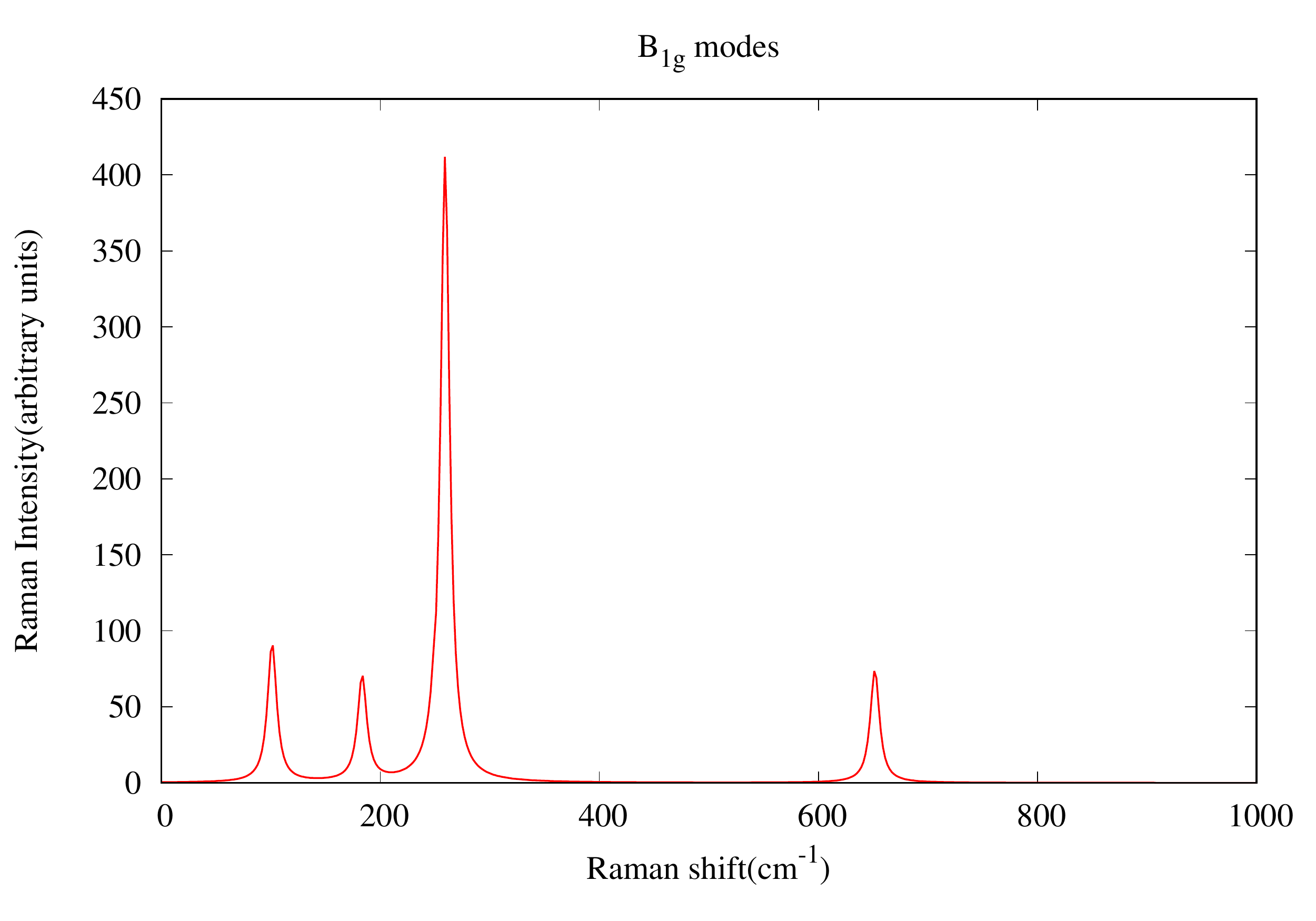}
 \caption{Raman spectrum for $B_{1g}$ modes, scattering geometry $z(xy)z$.\label{figramanb1g}}
\end{figure}

\begin{figure}
 \includegraphics[width=8cm]{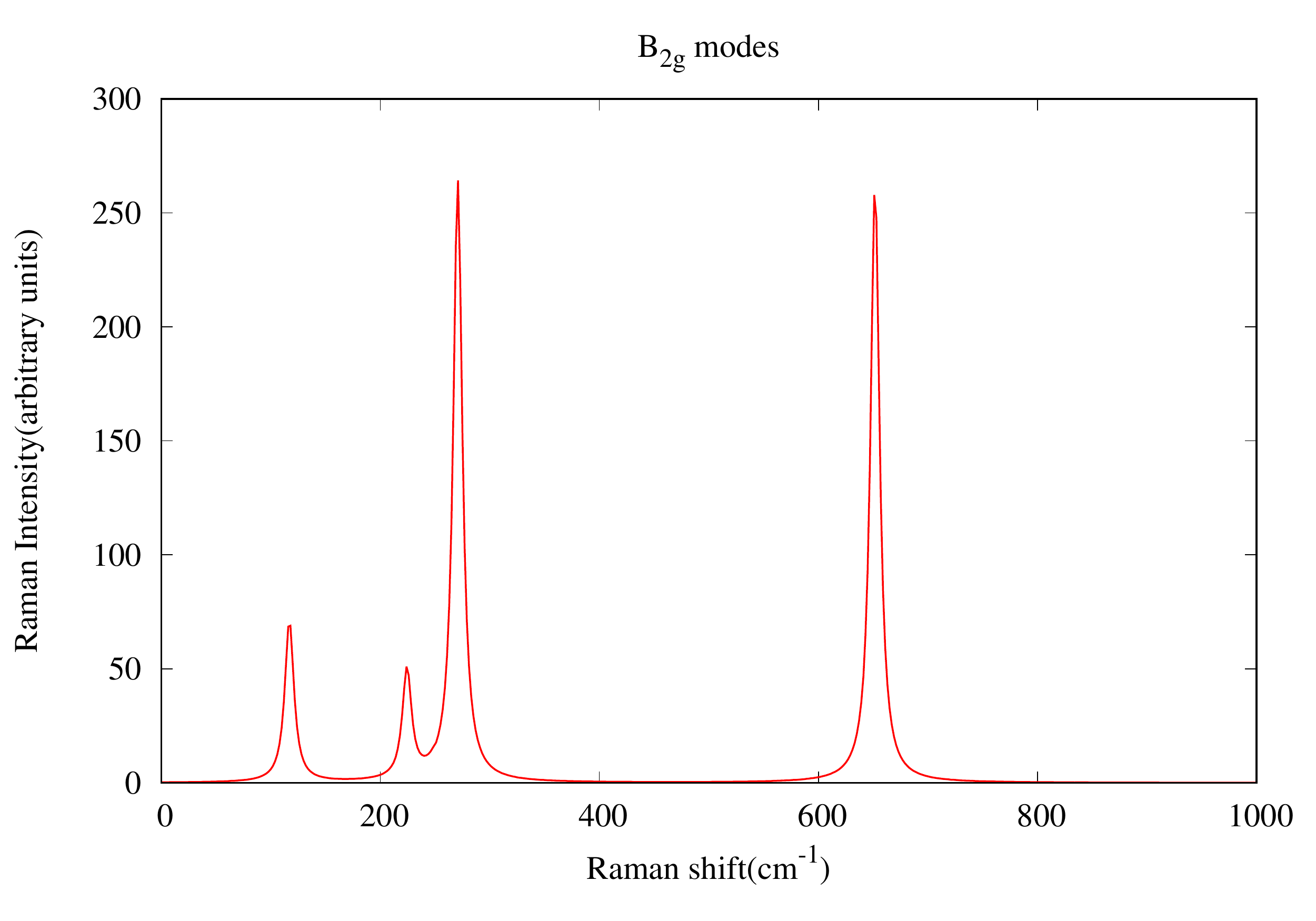}
 \caption{Raman spectrum for $B_{2g}$ modes, scattering geometry $y(xz)y$.\label{figramanb2g}}
\end{figure}

\begin{figure}
 \includegraphics[width=8cm]{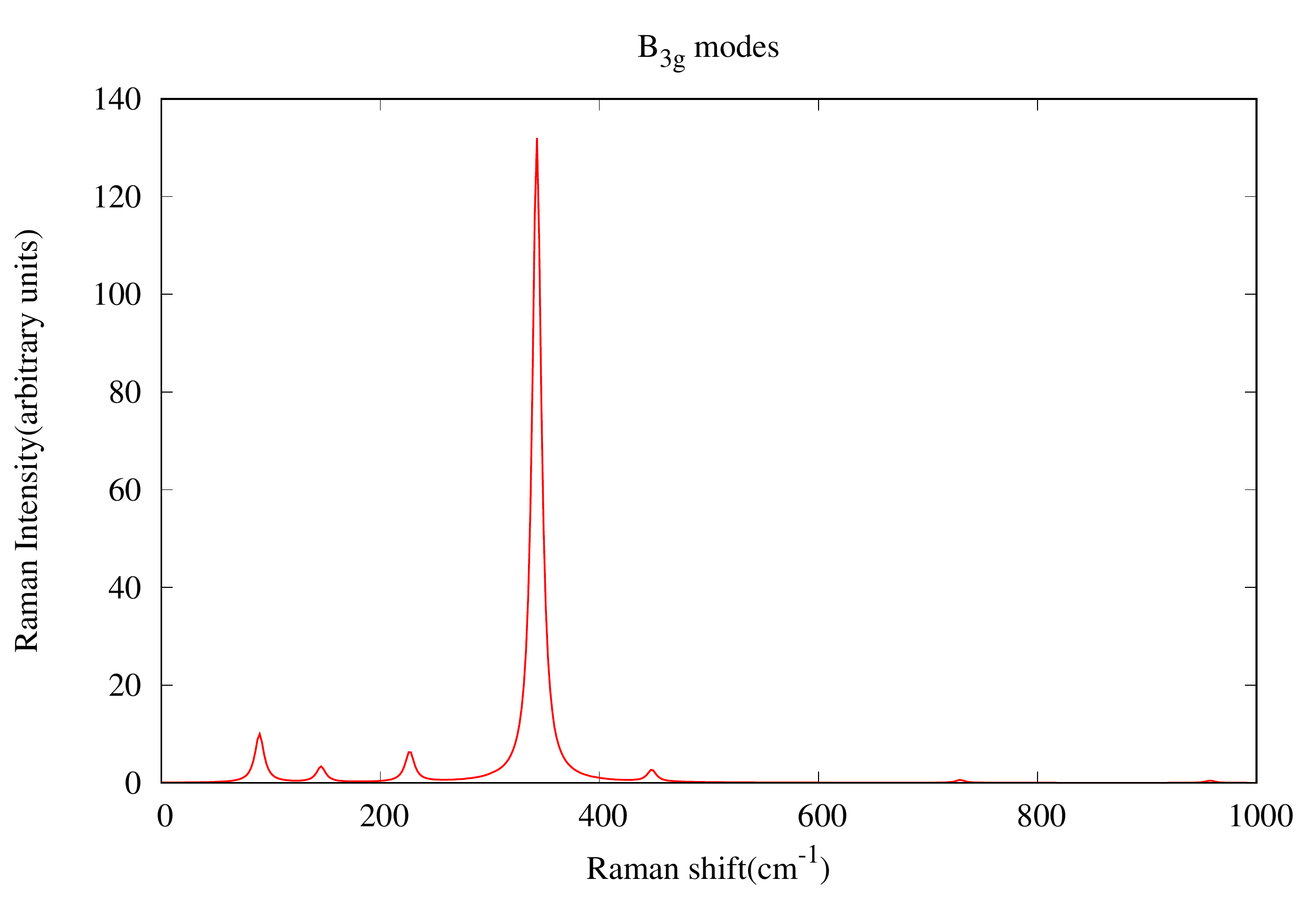}
 \caption{Raman spectrum for $B_{3g}$ modes, scattering geometry $x(yz)x$.\label{figramanb3g}}
\end{figure}

The results in this section were all obtained using the LDA calculations. 
The Raman tensor elements are given in Table \ref{tabraman}. The Raman
spectra for different scattering geometries, denoted
by ${\bf k}_i({\bf e}_i{\bf e}_o){\bf k}_o$ with
${\bf k}_{i/o}$ the incident/scattered  wavevector
and ${\bf e}_{i/o}$ the incident and scattered light polarization
are given in Figs. \ref{figramanag},\ref{figramanb1g}, \ref{figramanb2g}
and \ref{figramanb3g}.
For $A_g$ modes corresponding to parallel polarizations, the
intensity of the spectrum depends on the polarization selected.
For $z(xy)z$ (transmission) or $z(xy)\bar{z}$ (reflection) one measures $B_{1g}$ modes, for $xz$-polarizations one measures $B_{2g}$
and for $yz$ polarization one measures $B_{3g}$ modes.

One can see that the $A_g$ have by far the strongest intensities. The $B_{3g}$
modes are the weakest. The strongest $A_g$ mode at 727 cm$^{-1}$ in $yy$ polarization corresponds to a mode with mostly  in-plane eigendisplacements
of Mo-O$^{(1)}$ bond stretches. It also has fairly strong  $xx$ intensity
but negligible $zz$ motion because it does not involve motions normal
to the layer. On the other hand, the $A_g$ mode at 945 cm$^{-1}$ has its
strongest polarization as $zz$ and corresponds to  a Mo-O$^{(2)}$ stretch mode.
The strongest $B_{1g}$ mode is at 259 cm$^{-1}$ while the strongest $B_{2g}$
mode are at 270 cm$^{-1}$ and 652 cm$^{-1}$.
All modes below $\sim500$ cm$^{-1}$ are significantly weaker.
The three most prominent modes, $A_g$ at 945 cm$^{-1}$,  727 cm$^{-1}$
and the $B_{1g}$ mode at 652 cm$^{-1}$ agree well with the experimental
spectrum of Seguin \etal\cite{Seguin1995} apart from our
underestimates of these frequencies compared to the experiment. 

\subsection{Phonons in monolayer}

\begin{table*}
  \caption{Phonon frequencies in cm$^{-1}$ for monolayer compared to bulk $\alpha$-MoO$_3$ in PBEsol.
    M indicates monolayer, B bulk. For bulk an average of $\{A_g,B_{3g}\}$ is used for $A_g$,
    an average of $\{B_{1g},B_{2g}\}$ for $B_g$, $B_{3u}$ bulk corresponds to $A_u$ and
    an average of $\{B_{1u},B_{2u}\}$ for $B_u$.
 \label{tabmodemono}}
  \begin{ruledtabular}
    \begin{tabular}{dddddddddddd}
      \multicolumn{2}{c}{$A_g$} & \multicolumn{2}{c}{$B_g$} &  \multicolumn{2}{c}{$A_{uT}$} &\multicolumn{2}{c}{$A_{uL}$} &  \multicolumn{2}{c}{$B_{uT}$} &  \multicolumn{2}{c}{$B_{uL}$} \\ \hline
      \multicolumn{1}{c}{M}     &  \multicolumn{1}{c}{B} &  \multicolumn{1}{c}{M}     &  \multicolumn{1}{c}{B} &
      \multicolumn{1}{c}{M}     &  \multicolumn{1}{c}{B} & \multicolumn{1}{c}{M}     &  \multicolumn{1}{c}{B}  &
      \multicolumn{1}{c}{M}     &  \multicolumn{1}{c}{B} &  \multicolumn{1}{c}{M}     &  \multicolumn{1}{c}{B} \\ \hline
      75.0  &   73.7 &  91.7 &  100.2 & 179.0 & 178.9 &193.8 & 179.3 & 233.6 & 239.4 &  233.7 &  240.2 \\
      107.6 &  153.4 & 200.3 &  196.3 & 237.0 & 234.9 &342.2 & 336.3 & 336.5 & 339.2 &  339.3 &  343.5 \\
      218.4 &  222.3 & 258.8 &  264.6 & 457.2 & 477.6 &708.7 & 785.1 & 349.4 & 348.7 &  372.8 &  362.4 \\
      329.6 &  325.6 & 552.7 &  600.1 &       &       &      &       & 432.6 & 453.6 &  446.5 &  484.0 \\
      344.9 &  360.3 &       &        &       &       &      &       & 740.4 & 770.6 &  879.3 &  858.0 \\
      423.3 &  440.6 &       &        &       &       &      &       &1006.5 &1004.5 & 1006.8 & 1029.3 \\
      739.7 &  773.3 &       &        &       &       &      &       &       &       &        &        \\
      1011.7 & 1025.5&       &        &       &       &      &       &       &       &        &        \\
    \end{tabular}
  \end{ruledtabular}
\end{table*}

\begin{figure}
  \includegraphics[width=8cm]{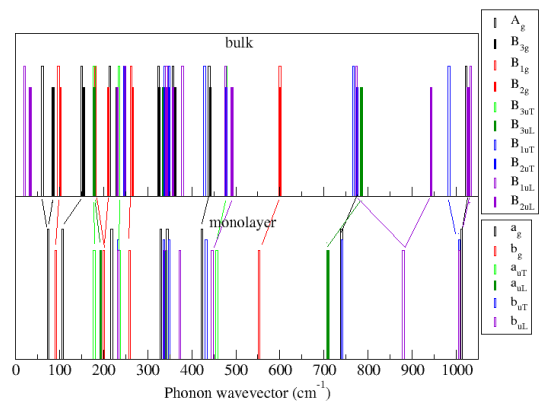}
  \caption{Comparison of bulk and monolayer phonon frequencies.\label{figbulkmono}}
\end{figure}

The phonons in a monolayer were calculated using the  PBEsol exchange-correlation
functional and using the PAW method. The results are shown in
Table \ref{tabmodemono}.
They are compared with the corresponding average of modes in the bulk calculated with the same functional.
As already mentioned in Table \ref{tabcharc2h}, there is a correspondence of irreducible representations
of the monolayer point group to those of the bulk orthorhombic structure point group. 

We note that we here treated the LO-TO splitting as in a 3D material.  Strictly speaking,
this is incorrect because in a monolayer, the LO-TO splitting goes
to zero  linearly in ${\bf q}_\parallel$ if we approach
the $\Gamma$-point closely enough. One should thus interpret the LO mode frequency here as being for
$|{\bf q}_\parallel|>1/r_{eff}$ with $r_{eff}$ an effective screening distance. 
This is due to the different nature of screening in a
2D material, which is unavoidably  wave-vector dependent.\cite{Sohier17}
It is given by $\varepsilon_{2D}({\bf q}_\parallel)=1+r_{eff}|{\bf q}_\parallel|)$, where
${\bf q}_\parallel$ is the wave vector parallel to the layer and $r_{eff}$ is an effective distance
approximately given by $r_{eff}\approx\varepsilon_\parallel^{bulk}d_M/2$ with $d_M$ the thickness of the 2D monolayer.\cite{Sohier17} Furthermore the Coulomb interaction in 2D has a different power dependence
on $|{\bf q}_\parallel|$, namely $2\pi/|{\bf q}_\parallel$ instead of $4\pi/|{\bf q}|^2$ in 3D. 

The correspondence with bulk mode and monolayer irreps was given in Table \ref{tabcharc2h}. 
The mode frequencies in bulk and monolayer are compared in Fig. \ref{figbulkmono}.
In the figure, we labeled monolayer irreps with lower case letters and color coded corresponding modes.
The height of the bars in this bar graph has no physical meaning but helps to distinguish close lying modes. 
The lowest modes of the bulk do not occur in the monolayer  because they
correspond to motions in which entire double layers
move with respect to each other. They are thus not included in Table \ref{tabmodemono} but
are shown in Fig. \ref{figbulkmono}. 
Several bulk (averages of corresponding modes) are seen to shift toward lower frequency
in the monolayer, for example the ($A_g^2$, $A_g^3$, $A_g^5$, $A_g^6$, $A_g^7$, $A_g^8$)
modes all have red-shifts. 
This is not too surprising since breaking even the weak bonds between layers would
reduce the stiffness of the system and hence
lead to smaller force constants and lower frequencies.

We can gain some more insights by looking specifically
at the high frequency modes, which correspond to a clear bond stretch.
For example the $A_g^8$, $B_{uT}^6$ modes both correspond to a Mo-O$^{(2)}$ bond stretch in the $z$ direction.
The bond stretch vibration  can be estimated as
\begin{equation}
  \omega\approx\sqrt{\frac{|K_{\mathrm{Mo}_z,\mathrm{O}^{(2)}_z}|}{\mu}}
\end{equation}
with
\begin{equation}
  \frac{1}{\mu}=\frac{1}{m_\mathrm{O}}+\frac{1}{m_{Mo}}
\end{equation}
the reduced mass if it were truly an isolated mode.
Using a frequency of about 1000 cm$^{-1}$ this corresponds to a force constant of
about 0.52 $e^2/a_0^3$ (or Hartree/Bohr$^2$).
Explicit calculations of the interatomic force constants show that
this force constant is $-0.41$ $e^2/a_0^3$. 
However part of this force has long-range dipolar character and changes in
screening may affect this dipolar part. Furthermore interactions across the van der Waals gap may
also affect these O$^{(2)}$-motion dominated modes. 
We here follow an analysis similar to that by Molina-S\'anchez and Wirtz\cite{MolinaSanchez11} for MoS$_2$ and
also used in Ref. \onlinecite{Bhandari14} for V$_2$O$_5$. 

First, we examine the changes in screening. 
The dielectric constant $\varepsilon^\infty$ of the monolayer
is given in Table \ref{tabdielmon} and compared with the corresponding bulk
as obtained within PBEsol. 
We can see that the dielectric constant in the plane is reduced
by about a factor 2 and
in the perpendicular direction by a factor 4.  In fact, this dielectric
constant is not really for a monolayer but
for a periodic system of monolayers spaced
by some large interlayer distance. For the $\varepsilon_{zz}^\infty$ we can
think of it as a capacitor of thickness $d$ filled with a layer of
thickness $d_M$ with dielectric constant $\varepsilon_M$ and vacuum in the rest.
The effective dielectric constant of the capacitor is then given
by $\varepsilon_{eff}^{-1}=1-\frac{d_M}{d}(1-\varepsilon_M^{-1})$.
In our calculation, the monolayer has thickness $d_M\approx 6.6$ \AA, while
the $c$ lattice constant is about $d=28.3$ \AA. This gives indeed an effective
dielectric constant of 1.2, which is  close to 1.44 in the actual calculation.
Strictly speaking, the in-plane dielectric constant should
become $\varepsilon_\parallel\approx1+\varepsilon_\parallel^{bulk}d_M|{\bf q}_\parallel|/2$, meaning that for ${\bf q}_\parallel\approx 1/d_M$  it is $1+\varepsilon_\parallel^{bulk}/2$ but for ${\bf q}_\parallel\rightarrow0$,
so at large distance, it will approach 1. 

\begin{table}
  \caption{Dielectric constant of monolayer and bulk system.\label{tabdielmon}}
\begin{ruledtabular}
  \begin{tabular}{lccc}
           &   $\varepsilon_{xx}^\infty$ & $\varepsilon_{yy}^\infty$ &  $\varepsilon_{zz}^\infty$ \\ \hline
 Monolayer &      3.01                     &  2.67                   &      1.44                 \\ 
 Bulk      &      5.95                     &  5.20                   &      4.00                  \\
  \end{tabular}
\end{ruledtabular}
\end{table}

Meanwhile it turns out that also the Born effective charges
change significantly.
For the monolayer calculation, they are given in Table \ref{tabbornmono}
as obtained within PBEsol. We here
give only the diagonal components.
We can see that the $Z^*_{zz}$ is reduced by almost a factor 3 for the monolayer,
while the in-plane components stay similar to the bulk.

Now, let's consider the vibrational modes corresponding to the Mo-O$^{(2)}$ bond
stretch.  The long-range dipolar force constant for this type of mode is
\begin{equation}
  K^{LR}_{\mathrm{Mo}_z,\mathrm{O}^{(2)}_z}=-2\frac{Z^*_{\mathrm{Mo}_{zz}}Z^*_{\mathrm{O}^{(2)}_{zz}}}{\sqrt{\varepsilon_{xx}\varepsilon_{yy}}d^3}
\end{equation}
where $d$ is the Mo-O$^{(2)}$ bond length. Clearly, because of the opposite sign Born charges, this interatomic force constant is positive.
This is opposite to what a short-range spring would do.
Indeed, if we move Mo in the $+z$ direction, the induced force
on the O$^{(2)}$ expected from a spring is also in the $+z$ direction
to counteract the compression of the spring. But the definition 
$K_{i\alpha,j\beta}=-\partial F_{j\beta}/\partial u_{i\alpha}$ then
implies a negative interatomic force constant.  
The strong Mo-O$^{(2)}$ bond implies that the total force constant is
negative and hence that the long-range part opposes the short range part.
This is similar to
the case of the V-O$^{vanadyl}$ in V$_2$O$_5$.\cite{Bhandari14}. 
We can see that
the Born charges here both are reduced by roughly a factor 3 while the denominator would be reduced by
a factor 2 from bulk to monolayer.  Hence the long range force constant is  reduced by a factor $\alpha=0.22$.
This dipolar part of the force constant from
the above equation amounts to $\sim$0.10 $e^2/a_0^3$
(as confirmed by explicit calculation to be $+0.0993$ $e^2/a_0^3$)
which is about 1/5 to 1/4 of the total force constant. The dipolar part being only a small part
thus is essentially quenched in the 2D system and
the force constant is reduced to only the short range part, which is larger. 
By itself this would then lead to an
increase in net force constant in the monolayer and a blue shift by about 8\%. 
However, the direct calculations shows a red-shift of smaller magnitude.  
This indicates that the above model of a localized isolated Mo-O$^{(2)}$ bond vibration is not sufficient.
We therefore surmise that the interaction between O$^{(2)}$ in adjacent layers
across the van der Waals gap must play a significant
role and act as an attractive force in the bulk system outweighing the screening change effect. It turns out the O$^{(2)}$ in one layer has interactions
with four neighboring O$^{(2)'}$  in the adjacent double layer. These
force constants are of order 0.0024$\pm0.0002$  $e^2/a_0$ in $x,y,z$
direction. Evaluating their effect on the frequency would require
a more detailed ball and spring model.  They actually also have a
substantial cancellation between long-range and short range components. 
It is clear however when these  forces are removed by increasing the
distance between the bilayers, then the frequency of the corresponding mode
will be  reduced.  In V$_2$O$_5$ \cite{Bhandari14}
the corresponding O$^{vanadyl}$-O$^{vanadyl\prime}$ was negligible because of
an almost exact compensation of the long-range
and short range parts. We may note from the structure, that the O$^{(2)}$ in
adjacent layers in MoO$_3$ are much closer together laterally (their interatomic distance is about 2.89 \AA)
than the vanadyl oxygens in V$_2$O$_5$ (interatomic distance 3.72 \AA).
Thus, the situation here is more similar to that in MoS$_2$ for the
out-of plane $A_g$ modes
as considered in Ref. \onlinecite{MolinaSanchez11}. We may note that the two effects considered
here oppose each other. 
The breaking of the weak interlayer interactions would lead  to a red shift and the change
in screening by itself would lead to a blue shift. Their compensation leads ultimately to
only a small shift of these modes. 


The modes $A_g^7$, $B_{uT}^5$ have a strong Mo-O$^{(1)}$ stretch character in the $y$ direction. This would
involve a dipolar force constant of the form
\begin{equation}
  K^{LR}_{\mathrm{Mo}_y,\mathrm{O}^{(1)}_y}=-2\frac{Z^*_{\mathrm{Mo}_{yy}}Z^*_{\mathrm{O}^{(1)}_{yy}}}{\sqrt{\varepsilon_{xx}\varepsilon_{zz}}d^3}
\end{equation}
In this case, one may notice that the Born charges barely change but the effective dielectric constant
in the denominator
is decreased by a factor 2.3. This would increase the dipolar part of
the force constant in the monolayer compared to the bulk.
The dipolar part in this case is again opposite to the short-range part
and even larger, the total force constant  $F_{yy}$ is found to be
$-0.270$  while the short range part is $-0.574$  and the long-range
part is $+0.304$  all in atomic units $e^2/a_0$. 
Thus, when the long-range part is increased by roughly a factor 2
the total force constant in this direction will be reduced in magnitude.
This in turn can explain the red-shift encountered by this mode.
This case differs from the MoS$_2$ case because there a Mo-Mo
force constant is in play which has the opposite sign and hence leads to a blue shift
for in-plane modes. It is similar to that case in the sense that the relevant Born charges
do not change appreciably but the dielectric constant does. 
On the other hand these modes also have a bond
stretch of the O$^{(2)}$ involved in their motion, so our analysis of 
the corresponding mode shift is here somewhat oversimplified. These
estimates are meant for the purpose of gaining insight only. 
We may expect from the analysis in the previous part
that this part of the motion in the $z$ direction is again influenced by the O$^{(2)}$-O$^{(2)\prime}$ interaction
between atoms in adjacent double layers. Thus overall, a stronger red-shift is expected and this is confirmed by
the direct calculations. It also agrees with the finding that this mode is particularly sensitive to the interlayer distance.

\begin{table}
  \caption{Born effective charges in monolayer compared with bulk calculated in PBEsol.\label{tabbornmono}}
  \begin{ruledtabular}
    \begin{tabular}{lddd}
      atom    & \multicolumn{1}{c}{$Z^*_{xx}$} &  \multicolumn{1}{c}{$Z^*_{yy}$} &  \multicolumn{1}{c}{$Z^*_{zz}$}  \\ \hline
              & \multicolumn{3}{c}{monolayer} \\ \hline 
      Mo        & 7.822     & 6.427 & 1.562 \\
      O$^{(1)}$     & -1.181  & -4.741 & -0.208 \\
      O$^{(2)}$     & -0.513  & -0.579 & -0.819 \\
      O$^{(3)}$     & -6.135  & -1.116 & -0.543 \\ \hline
      & \multicolumn{3}{c}{bulk} \\ \hline
      Mo    & 7.632 & 6.302 & 4.268 \\
      O$^{(1)}$ & -1.115 & -4.600 & -0.563 \\
      O$^{(2)}$ & -0.580 & -0.601 & -2.172 \\
      O$^{(3)}$ & -5.940 & -1.104 & -1.532 \\
    \end{tabular}
  \end{ruledtabular}
\end{table}

\section{Discussion and Conclusions}\label{conclusions}
In this paper we have presented a DFPT study of the phonons in
orthorhombic $\alpha$-MoO$_3$ with an emphasis on the Raman and infrared
spectra.
The calculated phonon frequencies  both in LDA and in PBEsol were
found to generally underestimate the experimental values slightly
but give comparable errors. The nature of the phonon spectrum
in terms of the eigenvectors was examined in some detail,
explaining why the modes occur in groups of four.

The intensities in Raman spectra and the assignments of the major
features are in good agreement with the experimental data reported in
Seguin \etal\cite{Seguin1995}, which also include previously
measured values.

Our paper also predicts shifts in some of the phonon frequencies in monolayer MoO$_3$
compared to bulk. The origin of these shifts was related to changes in
the dielectric screening and Born effective charges between bulk and monolayer but also
to the residual van der Waals interactions between the O$^{(2)}$ sticking out
from adjacent double layers.

While focusing on fundamental properties, our results may be anticipated to
be useful in future characterization of MoO$_3$ for applications which
require a thorough knowledge of the phonons. The polarization dependent Raman
spectra given here and details given in Supplemental Material on each of the phonon patterns may be particularly useful to investigate changes in some phonon
modes when hydrating the material or in some other way modifying the interlayer
distances. An overview of applications of MoO$_3$ can be found in
Ref. \onlinecite{deCastro17}. Furthermore the well separated
Restrahlen bands for different directions, with various ranges
in the mid-infrared where the index of refraction is negative in one
direction and positive in another provides opportunities for
natural hyperbolic materials and low-loss phonon-polaritons.\cite{Ma2018,Dixit2021}

{\bf Supplementary Material:}
Figures of the eigendisplacements of all modes are provided.

\acknowledgements{This work was supported by the Air Force Office of
  Scientific Research 
  under grant No. FA9550-18-1-0030. 
Calculations made use of the High Performance Computing Resource in the Core Facility for Advanced Research Computing at Case Western Reserve University.}
\bigskip
\par\noindent
{\bf Data Availability Statement:}\\
The data that supports the findings of this study are available within the article [and its supplementary material].\\

\par\noindent {\bf Conflict of Interest:}\\
  The authors have no conflicts to disclose.

\appendix
\section{Space group symmetries}
This appendix explains the space group symmetry operations in the $Pmcn$ setting
we use. The left two columns of Table \ref{space} correspond
to the setting of the International Tables of Crystallography (ITC).
The right two to the setting
used in our paper. The first column gives a short cut notation for
the symmetry operation, the second describes the symmetry operation
including its location
as follows: $2(00\frac{1}{2})(\frac{1}{4}0z)$ is the notation
used in ITC, meaning a 2 fold screw axis $2_{1z}$ along the $z$-axis
with $\frac{1}{2}$ translation along $z=c$ but located at $x=\frac{1}{4}$.
The second column gives the operation in form $\{{\bf R}|\vec{t}\}$
where ${\bf R}$ is a rotation matrix   and $\vec{t}$ is the non-primitive
translation. Note that because the screw axis is located a $1/4$ along $a$
it requires not only a translation along the symmetry
axis but also by 1/2 along $x$.
In our notation this becomes a 2-fold screw axis along $y$ with extra
translation along $z$ because it is located at $1/4$ along $c$.
Applied to the coordinates of a Wyckoff $4c$ site with coordinates
$(x,\frac{1}{4},z)$ this turns the atom in $(-x+\half,-\frac{1}{4},z+\half)$.
In our notation the $4c$ Wyckoff position is $(\frac{1}{4},y,z)$
and this operation turns it into $(-\frac{1}{4},y+\half,-z+\half)$.
where of course $-\frac{1}{4}$ can also be written $\frac{3}{4}$
because we can add any integer number to the fractional coordinates. 
Table \ref{tabwyck} shows how the 4 atoms of the Wyckoff site $4c$ transform
into each other and which symmetry operations relate them. 

\begin{table*}
  \caption{Space group elements in two settings. \label{space}}
  \begin{ruledtabular}
    \begin{tabular}{ccc|ccc}
      \multicolumn{3}{c}{Pnma} & \multicolumn{3}{c}{Pmcn} \\ 
      \multicolumn{3}{c}{$a>c>b$}&  \multicolumn{3}{c}{$c>b>a$} \\ \hline
      $1$      & $1$ &  $\left\{\left.\left(\begin{array}{ccc}1&&\\&1&\\&&1 \end{array}\right) \right| (0,0,0)\right\}$ &  $1$      & $1$ &  $\left\{\left.\left(\begin{array}{ccc}1&&\\&1&\\&&1 \end{array}\right) \right| (0,0,0)\right\}$\\
      $2_{1z}$ & $2(00\half)(\frac{1}{4}0z)$ & $\left\{\left.\left(\begin{array}{ccc}-1&&\\&-1&\\&&1 \end{array}\right) \right| (\half,0,\half)\right\}$ &
      $2_{1y}$ & $2(0\half0)(0 y \frac{1}{4})$ &  $\left\{\left.\left(\begin{array}{ccc}-1&&\\&1&\\&&-1\end{array}\right)\right|(0,\half,\half)\right\}$ \\
        $2_{1y}$ &   $2(0\half0)(0y0)$ & $\left\{\left.\left(\begin{array}{ccc}-1&&\\&1&\\&&-1 \end{array}\right) \right| (0,\half,0)\right\}$ & $2_{1x}$ &$2(\half00)(x00)$ & $\left\{\left.\left(\begin{array}{ccc}1&&\\&-1&\\&&-1\end{array}\right)\right|(\half,0,0)\right\}$ \\
          $2_{1z}$ &   $2(\half00)(x\quarter\quarter)$ & $\left\{\left.\left(\begin{array}{ccc}1&&\\&-1&\\&&-1 \end{array}\right) \right| (\half,\half,\half)\right\}$ & $2_{1z}$ &$2(00\half)(\quarter\quarter z)$ & $\left\{\left.\left(\begin{array}{ccc}-1&&\\&-1&\\&&1\end{array}\right)\right|(\half,\half,\half)\right\}$ \\               $-1$      & $-1 (000)$ &  $\left\{\left.\left(\begin{array}{ccc}-1&&\\&-1&\\&&-1 \end{array}\right) \right| (0,0,0)\right\}$ &  $-1$      & $-1 (000)$ &  $\left\{\left.\left(\begin{array}{ccc}-1&&\\&-1&\\&&-1 \end{array}\right) \right| (0,0,0)\right\}$\\
            $m_y$ & $m(x\quarter z)$ &  $\left\{\left.\left(\begin{array}{ccc}1&&\\&-1&\\&&1 \end{array}\right) \right| (0,\half,0)\right\}$ & $m_x$ & $m(\quarter yz)$ &  $\left\{\left.\left(\begin{array}{ccc}-1&&\\&1&\\&&1 \end{array}\right) \right| (\half,0,0)\right\}$ \\
            $a_z$ & $a(xy\quarter)$ &  $\left\{\left.\left(\begin{array}{ccc}1&&\\&1&\\&&-1 \end{array}\right) \right| (\half,0,\half)\right\}$ &
            $c_y$ & $c(x\quarter z)$ & $\left\{\left.\left(\begin{array}{ccc}1&&\\&-1&\\&&1 \end{array}\right) \right| (0,\half,\half)\right\}$ \\
            $n_x$ & $n(0\half\half)(\quarter yz)$ &  $\left\{\left.\left(\begin{array}{ccc}-1&&\\&1&\\&&1 \end{array}\right) \right| (\half,\half,\half)\right\}$ &
            $n_z$ & $n(\half\half0)(xy\quarter)$ &  $\left\{\left.\left(\begin{array}{ccc}1&&\\&1&\\&&-1 \end{array}\right) \right| (\half,\half,\half)\right\}$
    \end{tabular}
  \end{ruledtabular}
\end{table*}

\begin{table}
  \caption{Transformation rule for Wyckoff position $4c$ in the space group
    $Pmcn$ setting. 
    The last column lists the symmetry elements transforming
    the first atom to the one in that row.\label{tabwyck}}
  \begin{ruledtabular}
    \begin{tabular}{cccc}
      $\quarter$ & $y$ & $z$ & $1$,$m_x$ \\
      $\frac{3}{4}$ & $-y$ & $-z$& $2_{1x}$,$-1$ \\
      $\quarter$    & $-y+\half$ & $z+\half$ & $2_{1z}$, $c_y$ \\
      $\frac{3}{4}$ & $y+\half$ &$-z+\half$ & $2_{1y}$,$n_z$ \\
    \end{tabular}
  \end{ruledtabular}
\end{table}

\bibliography{moo3,dft,abinit}
\end{document}